\documentclass[10pt]{amsart}
\usepackage[small]{caption}
\usepackage{amsmath,amsfonts,amscd,amssymb,epsf, epsfig}\usepackage{graphicx}

\theoremstyle{definition}

\theoremstyle{remark} 
\numberwithin{equation}{section}

\newcommand{\Z}{{\mathbb{Z}}}

\newcommand{\C}{\mathbb{C}}

\newcommand{\R}{\mathbb{R}}

\newcommand{\pa}{\partial}

\newcommand{\vep}{\varepsilon}

\begin{document}

\title{Fractional Oscillator Process with two indices}
\author{S.C. Lim}\address{ Faculty of Engineering,
Multimedia University, Jalan Multimedia, Cyberjaya, 63100, Selangor
Darul Ehsan, Malaysia.} \email{sclim@mmu.edu.my}

\author{L.P. Teo}\address{Faculty of Information
Technology, Multimedia University, Jalan Multimedia, Cyberjaya,
63100, Selangor Darul Ehsan, Malaysia.}\email{lpteo@mmu.edu.my}

 \maketitle

\begin{abstract}
We introduce a new fractional oscillator process which can be
obtained as solution of  a stochastic differential equation with two
fractional orders. Basic properties such as fractal dimension and
short range dependence of the process are studied by considering the
asymptotic properties of its covariance function. The
fluctuation--dissipation relation of the process is investigated.
The fractional oscillator process can be regarded as one-dimensional
fractional Euclidean Klein-Gordon field, which can be obtained by
applying the Parisi-Wu stochastic quantization method to a nonlocal
Euclidean action. The Casimir energy associated with the fractional
field at positive temperature is calculated by using the zeta
function regularization technique.

\end{abstract}

\section{Introduction}
Work on extending stochastic processes characterized by a single
index to corresponding processes with two indices or processes with
variable index has attracted considerable interest lately. For
example, fractional Brownian motion parametrized by a constant Hurst
index $H$ has recently been generalized to the bifractional Brownian
motion \cite{nr1, nr2} characterized by two indices, with fractional
Brownian motion as a special case.  A process with two indices
provides a more flexible model as far as applications are concerned.
Fractional Brownian motion has also been extended to multifractional
Brownian motion with a variable index $H(t)$ \cite{nr3,nr4}.
Multifractional Brownian motion allows one to model phenomena with
variable memory and fractal dimension that changes with time or
position. Another example is the process of Cauchy class \cite{nr5},
which has been extended to the process with two-indices known as
generalized Cauchy process \cite{nr6, nr7}. In contrast to many
other single-index processes (for example, fractional Brownian
motion), generalized Cauchy process has the advantage that its two
indices separately characterize the fractal dimension or
self-similar property and the long range dependence. The aim of this
paper is to introduce a new type of fractional Gaussian process
indexed by two parameters so that its short range dependence
property and fractal dimension can be separately characterized. Such
a process can be regarded as a generalization of fractional
oscillator process or fractional Ornstein-Uhlenbeck process with
single index \cite{nr8, nr9}.

    This paper is organized as follows. Section 2 introduces the
    generalized fractional oscillator process as solution of a
stochastic differential equation with two fractional orders. Despite
the covariance of this process does not have a closed analytic form,
its  basic properties can be studied by considering the asymptotic
properties of its covariance. In particular, the fractal dimension
and short range dependence are studied. The fluctuation--dissipation
relation are discussed in section 4. In  section 5, we consider the
fractional oscillator process as one-dimensional Euclidean
fractional scalar field. Stochastic quantization of the field at
zero and finite temperature is carried out. We proceed to calculate
the Casimir energy associated with the fractional field at finite
temperature by employing the zeta function regularization technique.
In the last section, we briefly discuss possible generalizations of
the results obtained.

\section{Fractional Oscillator Process with two indices}

In this section we define the fractional oscillator process with two
indices.  Recall that an ordinary oscillator process $X(t)$ can be
obtained as a solution to the Langevin equation
\begin{align}\label{eq3_19_1}
\left(D_t+\lambda\right)X(t)=\eta(t),
\end{align}where $\eta(t)=dB(t)/dt$ is the standard white noise with
\begin{align*}
\langle \eta(t)\rangle =0, \hspace{1cm}\langle
\eta(t)\eta(t')\rangle =\delta(t-t'),
\end{align*}$B(t)$ is the standard Brownian motion, and $\lambda$ is a positive constant. Using
Fourier transform, the solution of \eqref{eq3_19_1} can be written
as
\begin{align*}
X(t)=\frac{1}{\sqrt{2\pi}}\int\limits_{\R}\frac{e^{i\omega
t}\hat{\eta}(\omega)}{i\omega+\lambda}d\omega,
\end{align*}
where $\hat{\eta}(\omega)$ is the Fourier transform of the standard
white noise $\eta(t)$:
\begin{align*}
\hat{\eta}(\omega)=\frac{1}{\sqrt{2\pi}}\int_{-\infty}^{\infty}e^{-it\omega}dB(t).
\end{align*}In the literature, $X(t)$ is known as the oscillator process or the Ornstein--Uhlenbeck process.
It  is a centered stationary Gaussian process with covariance
function
\begin{align}\label{eq3_19_2}
\langle X(s)X(s+t)\rangle =
\frac{1}{2\pi}\int\limits_{\R}\frac{e^{i\omega
t}}{\omega^2+\lambda^2}d\omega =\frac{e^{-\lambda|t|}}{2\lambda}.
\end{align} One can also regard the oscillator process $X(t)$ as
one-dimensional Euclidean scalar Klein--Gordon field with mass
$\lambda$, and propagator given by the spectral density
$$S(\omega)=\frac{1}{2\pi}\frac{1}{\omega^2+\lambda^2}.$$

Since fractal dynamics \cite{A1, A2, A3} have increasingly played an
important role in various transport phenomena in complex media, it
would be interesting to investigate various possible generalizations
of $X(t)$   to its fractional counterpart. One direct way is to
replace the differential operator $D_t$  in \eqref{eq3_19_1} by the
fractional differential operator $_aD_t$ to obtain the following
type I fractional Langevin equation \cite{new2, new3, new4, new5}:
\begin{align*}
\left(_aD_t^{\alpha}+\lambda\right)\mathsf{X}_{\alpha,1}(t)=\eta(t),
\end{align*}where  the fractional derivative $_aD_t$  is defined as
 \cite{nr10, IP, nr12}:
\begin{align*}
\left(\,_aD_t^{\alpha}f\right)(t)=\frac{1}{\Gamma(n-\alpha)}\left(\frac{d}{dt}\right)^n\int_a^{t}
\frac{f(u)}{(t-u)^{\alpha-n+1}}du,\hspace{1cm} n-1<\alpha<n.
\end{align*}
When  $a=0$, $_0D_t^{\alpha}$   is known as the Riemann--Liouville
fractional derivative; and for $a=-\infty$, $_{-\infty}D_t^{\alpha}$
is called the Weyl fractional derivative. Another possible
generalization is to fractionalize the operator $(_aD_t+\lambda)$ to
obtain the following type II fractional Langevin equation \cite{nr8,
nr9}:\begin{align*}
\left(_aD_t+\lambda\right)^{\gamma}\mathsf{X}_{1,\gamma}(t)=\eta(t),\hspace{1cm}\gamma>0.
\end{align*}Recently we have combined the investigation on these two
type of processes and study the more general case \cite{new1}:
\begin{align}\label{eq4_22_1}
\left(_aD_t^{\alpha}+\lambda\right)^{\gamma}\mathsf{X}_{\alpha,\gamma}(t)=\eta(t).
\end{align} If Weyl fractional derivative is used in \eqref{eq4_22_1}, then for $\alpha\gamma>1/2$,
$\mathsf{X}_{\alpha,\gamma}(t)$ turns out to be a centered
stationary Gaussian process with a representation
\begin{align*}
\mathsf{X}_{\alpha,\gamma}(t)=\frac{1}{\sqrt{2\pi}}\int\limits_{\R}\frac{e^{i\omega
t}\hat{\eta}(\omega)}{\left((i\omega)^{\alpha}+\lambda\right)^{\gamma}}d\omega,
\end{align*}and covariance function
\begin{align}\label{eq3_19_4}
\langle
\mathsf{X}_{\alpha,\gamma}(s)\mathsf{X}_{\alpha,\gamma}(s+t)\rangle
=\frac{1}{2\pi}\int\limits_{\R}\frac{e^{i\omega
t}}{\left|(i\omega)^{\alpha}+\lambda\right|^{2\gamma}}d\omega=\frac{1}{2\pi}\int\limits_{\R}
\frac{e^{i\omega t}}{\left(|\omega|^{2\alpha}+2\lambda
|\omega|\cos\frac{\pi\alpha}{2}+\lambda^2\right)^{\gamma}}d\omega.
\end{align}The properties of the process $\mathsf{X}_{\alpha,\gamma}(t)$
have been studied in \cite{new1}. However, the above generalization
contains an unsatisfactory aspect, namely  its spectral density
\begin{align}\label{eq3_19_5}S(\omega)=\frac{1}{2\pi}\frac{1}{\left(|\omega|^{2\alpha}+2\lambda
|\omega|\cos\frac{\pi\alpha}{2}+\lambda^2\right)^{\gamma}}\end{align}
has a complicated form. When $\alpha=1$, the spectral density
simplifies to
\begin{align*}
S(\omega)=\frac{1}{2\pi}\frac{1}{\left(|\omega|^{2}+\lambda^2\right)^{\gamma}}.
\end{align*}This signifies that $\mathsf{X}_{1,\gamma}(t)$ has another representation
\begin{align*}
\mathsf{X}_{1,\gamma}(t)=\frac{1}{\sqrt{2\pi}}\int\limits_{\R}\frac{e^{i\omega
t}}{\left(|\omega|^{2}+\lambda^2\right)^{\frac{\gamma}{2}}}d\omega,
\end{align*}and therefore is  a solution of the following fractional stochastic differential equation
\begin{align*}
\left(-\Delta+\lambda^2\right)^{\frac{\gamma}{2}}\mathsf{X}_{1,\gamma}(t)=\eta(t),
\end{align*}where $\Delta=d^2/dt^2$ is the one-dimensional Laplacian
operator. Inspired by this, we define a new type of stochastic
process $X_{\alpha,\gamma}(t)$ indexed by two parameters
$\alpha,\gamma$ with $\alpha\in (0,1]$, $\gamma>0$, as a solution to
the following fractional stochastic differential equation:
\begin{align*}
\left[\left(-\Delta\right)^{\alpha}+\lambda^2\right]^{\frac{\gamma}{2}}X_{\alpha,\gamma}(t)
=\eta(t),
\end{align*}or equivalently, the equation
\begin{align}\label{eq3_23_1}
\left(\mathbf{D}_t^{2\alpha}+\lambda^2\right)^{\frac{\gamma}{2}}X_{\alpha,\gamma}(t)=\eta(t),
\end{align}where $\mathbf{D}_t^{2\alpha}$ is the Riesz derivative
defined by \cite{nr10, nr13} \begin{align*} \mathbf{D}_t^{2\alpha}
f:=(-\Delta)^{\alpha}f:=F^{-1}\left(|\omega|^{2\alpha}\hat{f}(\omega)\right),
\end{align*}with $\hat{f}:=F(f)$ the Fourier transform of $f$. It follows easily that when
$\alpha\gamma>1/2$, the solution to \eqref{eq3_23_1} is given by
\begin{align}\label{eq3_21_1}
X_{\alpha,\gamma}(t)=\frac{1}{\sqrt{2\pi}}\int\limits_{\R}\frac{e^{i\omega
t}}{\left(|\omega|^{2\alpha}+\lambda^2\right)^{\frac{\gamma}{2}}}d\omega.
\end{align}We also call $X_{\alpha,\gamma}(t)$ a fractional
oscillator process. Its covariance function
$C_{\alpha,\gamma}(t)=\langle
X_{\alpha,\gamma}(s+t)X_{\alpha,\gamma}(s)\rangle$  is
\begin{align}\label{eq3_21_2}
C_{\alpha,\gamma}(t)=\frac{1}{2\pi}
\int_{-\infty}^{\infty}\frac{e^{it\omega}}{\left(|\omega|^{2\alpha}+\lambda^2\right)^{\gamma}}d\omega,
\end{align} and the spectral density has the required simpler form
\begin{align}\label{eq3_14_2}
S_{\alpha,\gamma}(\omega)=
\frac{1}{2\pi}\frac{1}{\left(|\omega|^{2\alpha}+\lambda^2\right)^{\gamma}}
\end{align}compared to \eqref{eq3_19_5}. Some simulations of the
 process $X_{\alpha,\gamma}(t)$ are given in
Figure 1.
\begin{figure}\centering \epsfxsize=.49\linewidth
\epsffile{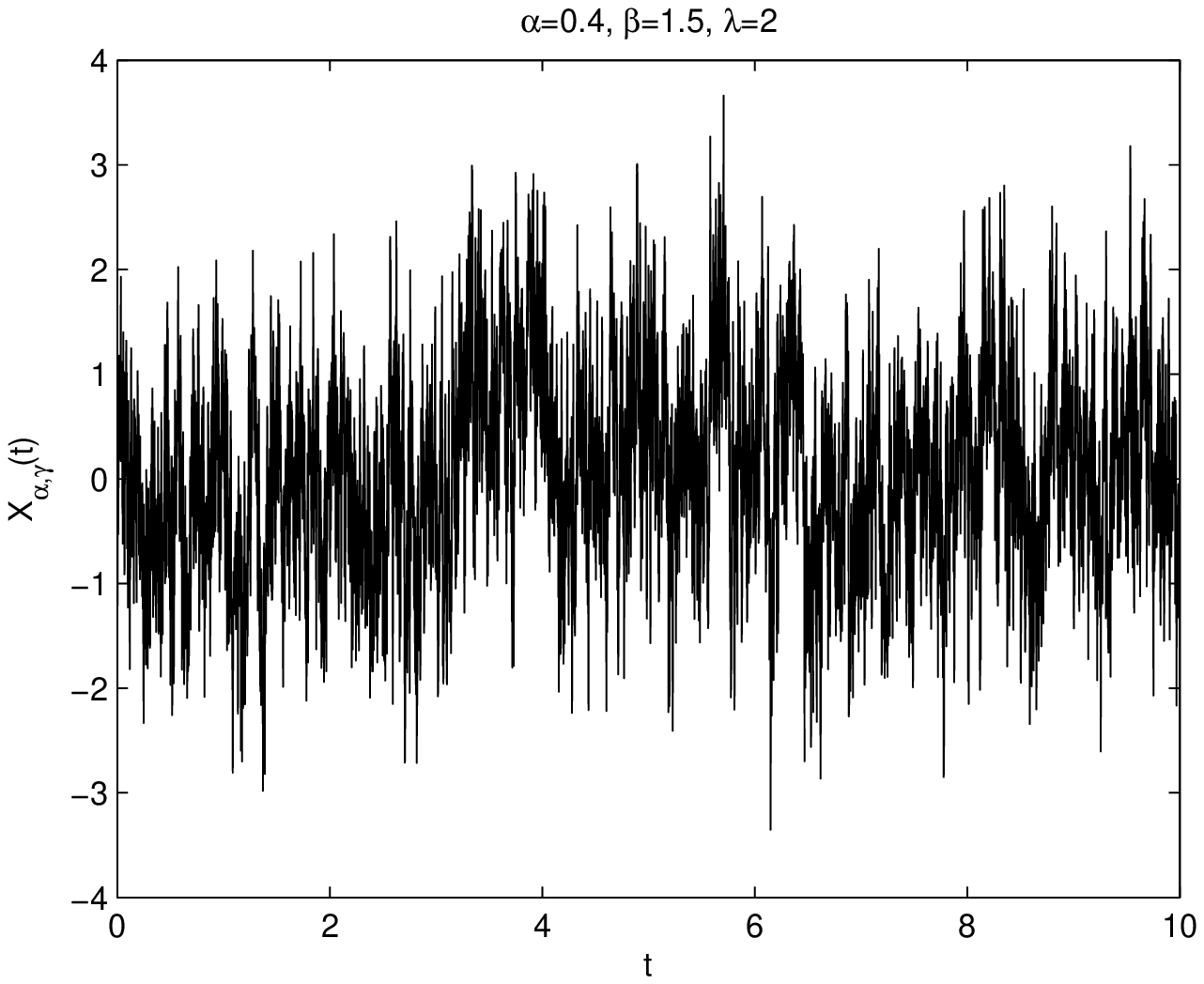}\centering \epsfxsize=.49\linewidth
\epsffile{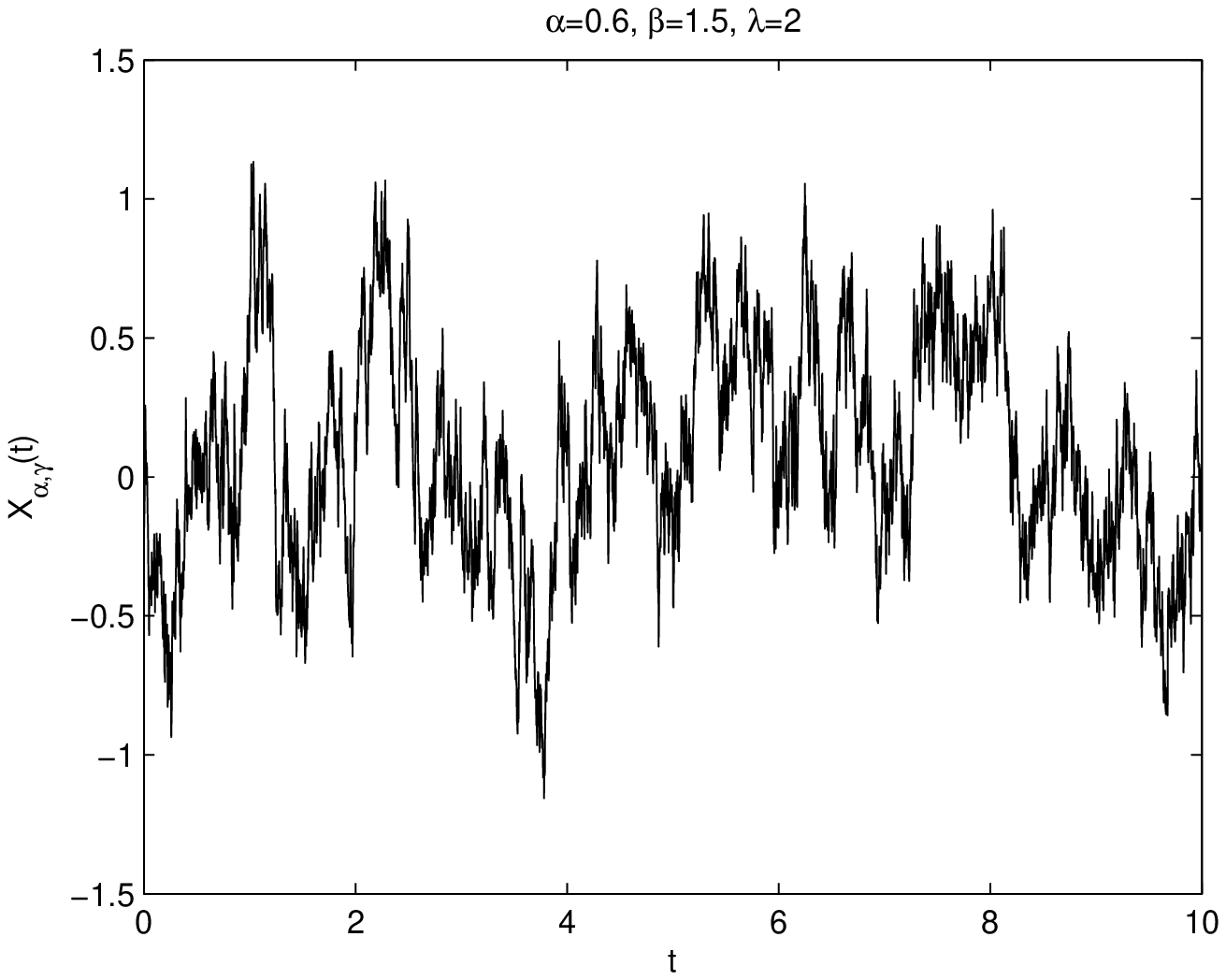}\\
\centering \epsfxsize=.49\linewidth \epsffile{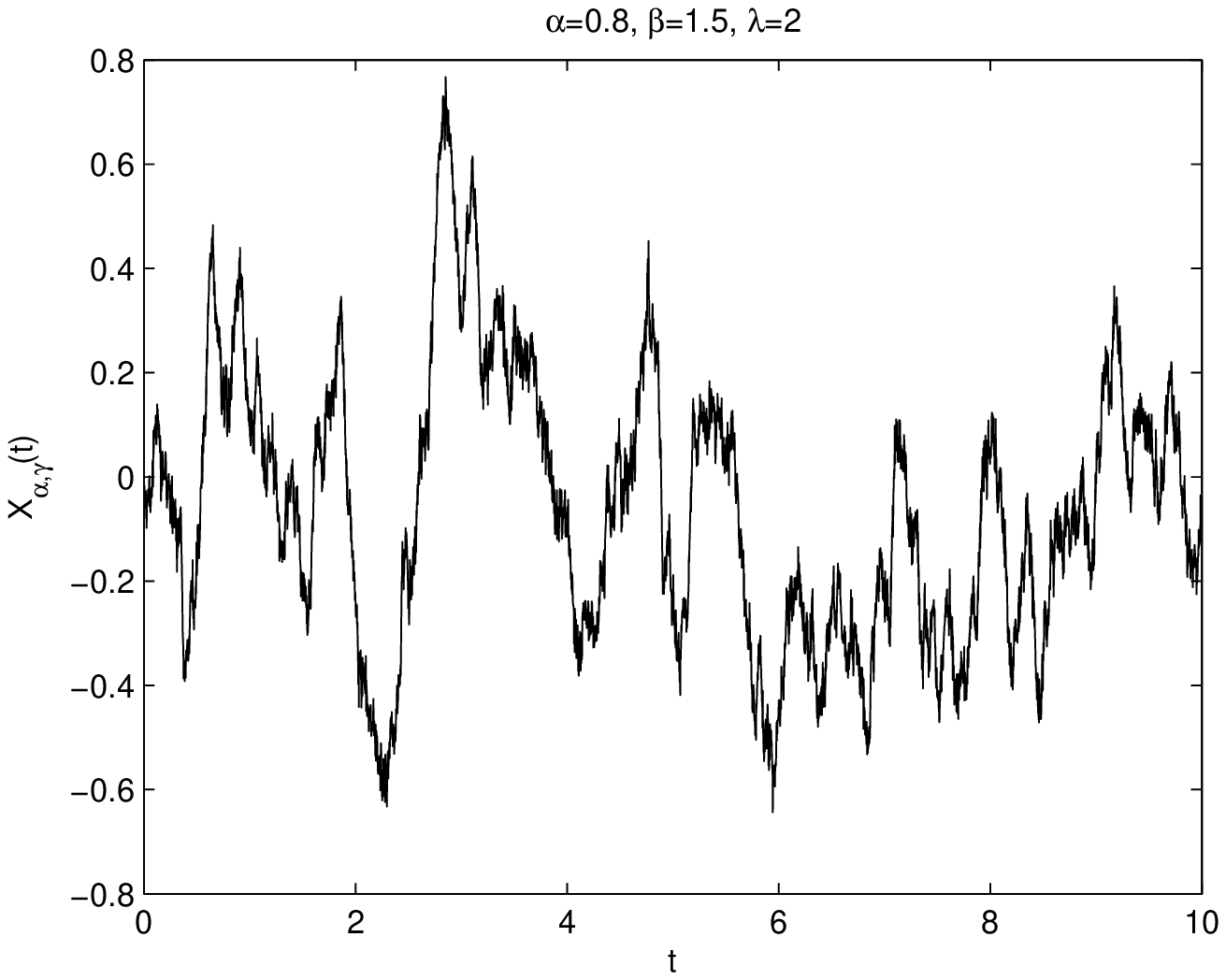} \centering
\epsfxsize=.49\linewidth \epsffile{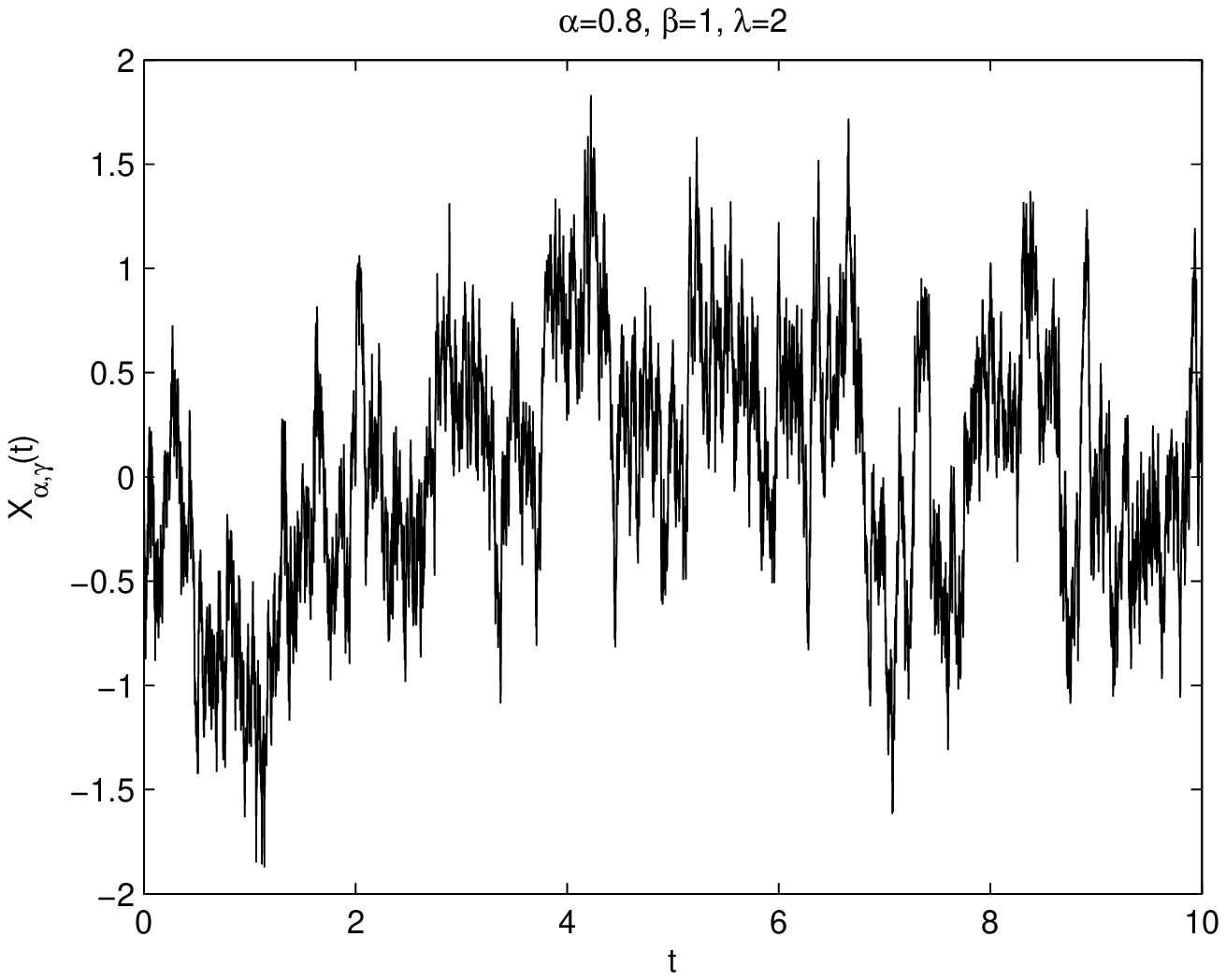}\caption{Some
simulations of $X_{\alpha,\gamma}(t)$ for different values of
$\alpha$ and $\gamma$.}\end{figure}

Here we would also like to remark that when $\alpha\gamma\leq 1/2$,
$X_{\alpha,\gamma}(t)$ only exists as a generalized stochastic
process, in the sense of generalized functions. Namely, for a
Schwarz class test function $f(t)$,
\begin{align*}
\left\langle X_{\alpha,\gamma}(t), f(t)\right\rangle =
\int\limits_{\R}
\frac{\hat{f}(-\omega)\hat{\eta}(\omega)}{\left(|\omega|^{2\alpha}+\lambda^2\right)^{\frac{\gamma}{2}}}d\omega,
\end{align*}where $\hat{f}(\omega)$ is the Fourier transform of $f$.

The two fractional generalizations $\mathsf{X}_{\alpha,\gamma}(t)$
and $X_{\alpha,\gamma}(t)$ of ordinary oscillator process do not
give the same solutions since they give rise to different spectral
densities. However, if we want to regard fractional oscillator
process as one-dimensional Euclidean Klein--Gordon field,
\eqref{eq3_23_1} is one-dimensional Klein--Gordon equation with two
fractional orders, and the covariance function \eqref{eq3_21_2}
becomes the propagator of the corresponding fractional Klein-Gordon
field $\phi_{\alpha,\gamma}(t)$. We shall show in section 5 that
$\phi_{\alpha,\gamma}(t)$ can be obtained by applying Parisi-Wu
stochastic quantization \cite{nr14} involving a nonlocal Euclidean
action.

\section{Asymptotic behaviors of the covariance function
$C_{\alpha,\gamma}(t)$ and sample path properties of
$X_{\alpha,\gamma}(t)$}

\subsection{Asymptotic behaviors of the covariance function
$C_{\alpha,\gamma}(t)$.}
   When
    $\alpha=1$,
$\gamma>0$, the covariance function $C_{1,\gamma}(t)$
\eqref{eq3_21_2} has the following closed form \cite{nr8}:
\begin{align}\label{eq8}  C_{1,\gamma}(t)=
 \frac{2^{1/2-\gamma}}{\sqrt{\pi} \Gamma(\gamma)}\left(\frac{|t|}{\lambda}\right)
^{\gamma-1/2}K_{\gamma-1/2}(\lambda|t|),
\end{align}where $K_{\nu}(z)$ is the modified Bessel function of
second kind or the MacDonald function. However, the covariance
function $C_{\alpha,\gamma}(t)$ in general does not exist in closed
analytic form. In fact, since the spectral density \eqref{eq3_14_2}
has the same functional form as the characteristic function of the
generalized Linnik distribution \cite{14}, the covariance function
$C_{\alpha,\gamma}(t)$ \eqref{eq3_21_2} has the same functional form
as the Linnik probability density function, whose analytic
properties have been studied in \cite{17, 14}. In particular, one
can obtain the following integral representation for the covariance
function $C_{\alpha,\gamma}(t)$:
\begin{align}\label{eq12_21_2}
C_{\alpha,\gamma}(t)=\frac{1}{\pi}\text{Im}\int_0^{\infty}\frac{e^{-u|t|}du}{\left(\lambda^2+e^{-i\pi\alpha}u^{2\alpha}
\right)^{\gamma}}.
\end{align} It turns out that the
analytic properties of Linnik probability density function depend on
the arithmetic nature of the parameters $\alpha$ and $\gamma$; and
the conditions imposed on $\alpha$ and $\gamma$ are rather
complicated and are not of practical interest. Therefore, we shall
use different methods to study the asymptotic behaviors of the
covariance function $C_{\alpha,\gamma}(t)$ that are more suited for
various applications.

For the properties of $X_{\alpha,\gamma}(t)$  that we are interested
such as its fractal dimension, long or short range dependence, it
suffices for us to know  the leading behavior of the variance of the
associated increment process
\begin{align}\label{eq3_23_3}\sigma^2(t)=\left\langle\left[X_{\alpha,\gamma}
(s+t)-X_{\alpha,\gamma} (s)\right]^2\right\rangle
=2C_{\alpha,\gamma} (0)-2C_{\alpha,\gamma} (t)\end{align} for $t
\rightarrow 0$ and the leading behavior of $C_{\alpha,\gamma}(t)$
for $t\rightarrow \infty$.

    First, we examine the behavior of   $\sigma^2(t)$ when $t\rightarrow 0$.
We impose the restriction $\alpha\gamma>1/2$ so that
$X_{\alpha,\gamma}(t)$ has finite variance and \eqref{eq3_23_3} is
well-defined. Under this restriction, the variance
$C_{\alpha,\gamma}(0)$  is given by (\#3.251, no.11, Ref. \cite{15})
\begin{align}\label{eq4_16_1}
C_{\alpha,\gamma}(0)=&\frac{1}{\pi}\int_0^{\infty}
\frac{d\omega}{(\omega^{2\alpha}+\lambda^2)^{\gamma}}\\
=& \frac{1}{2\pi \alpha}\frac{\Gamma\left(\frac{1}{2\alpha}\right)
\Gamma\left(\gamma-\frac{1}{2\alpha}\right)}{\Gamma(\gamma)}\lambda^{\frac{1}{\alpha}-2\gamma}\nonumber\\
=&\frac{\lambda^{\frac{1}{\alpha}-2\gamma}}{2\alpha\Gamma(\gamma)}
\frac{\Gamma\left(\frac{1}{2\alpha}\right)}{
\Gamma\left(1-\gamma+\frac{1}{2\alpha}\right)\sin\left(\pi\left(\gamma-\frac{1}{2\alpha}\right)\right)},\nonumber\end{align}
where the identity $-z\Gamma(z)\Gamma(-z)=\pi/\sin(\pi z)$ has been
used. Our result is in agreement with that of Erdogan and Ostrovskii
\cite{14}. To obtain the leading behavior of
\begin{align}\label{eq12_21_3}\sigma^2(t)=2C_{\alpha,\gamma}(0)-2C_{\alpha,\gamma}(t)=\frac{2}{\pi}\int_0^{\infty}
\frac{1-\cos(\omega|t|)}{(\lambda^2+\omega^{2\alpha})^{\gamma}}d\omega,\end{align}
we consider the cases  $1/2<\alpha\gamma<3/2$, $\alpha\gamma>3/2$
and $\alpha\gamma=3/2$ separately.

\noindent Case I. When $1/2<\alpha\gamma<3/2$,  \eqref{eq12_21_3} is
equal to
\begin{align}\label{eq12_24_5}
\sigma^2(t)=&\frac{4}{\pi}\int_0^{\infty}
\frac{\sin^2\left(\frac{\omega|t|}{2}\right)}{(\lambda^2+\omega^{2\alpha})^{\gamma}}d\omega
\\=&\frac{4|t|^{2\alpha\gamma-1}}{\pi}\int_0^{\infty}
\frac{\sin^2(\omega/2)}{(\omega^{2\alpha}+\lambda^2|t|^{2\alpha})^{\gamma}}d\omega\nonumber\\
= &
\frac{4|t|^{2\alpha\gamma-1}}{\pi}\int_0^{\infty}\omega^{-2\alpha\gamma}\sin^2\left(\frac{\omega}{2}\right)d\omega
+o(|t|^{2\alpha\gamma-1})\nonumber\\=&-
\frac{|t|^{2\alpha\gamma-1}}{\cos(\pi\alpha\gamma)\Gamma(2\alpha\gamma)}+o(|t|^{2\alpha\gamma-1})\hspace{1cm}\text{as}\;\;t\rightarrow
0.\nonumber
\end{align}We have used \#3.823 of \cite{15} in the last step. Eq.
\eqref{eq12_24_5} shows that when $1/2<\alpha\gamma<3/2$ and
$t\rightarrow 0$, the leading term of $\sigma^2(t)$ is of order
$|t|^{2\alpha\gamma-1}$. If we replace $\alpha\gamma$ by $H+1/2$,
then \eqref{eq12_24_5} becomes
\begin{align*}
\sigma^2(t)\sim \frac{|t|^{2H}}{\sin (\pi
H)\Gamma(2H+1)}+o(|t|^{2H})\hspace{1cm}\text{as}\;\; t\rightarrow 0,
\end{align*}which shows that the short time asymptotic behavior of
$\sigma^2(t)$ is characterized by the index $H= \alpha\gamma-1/2$.

\noindent Case II. For $\alpha\gamma>3/2$, using $1-\cos(\omega|t|)
=\frac{1}{2}\omega^2|t|^2+O(|t|^4)$ as $t\rightarrow 0$, we have
\begin{align}\label{eq12_24_4}
\sigma^2(t)=&\frac{|t|^2}{\pi}\int_0^{\infty}\frac{\omega^2
d\omega}{(\lambda^2+\omega^{2\alpha})^{\gamma}}+o(|t|^2)\\
=&\frac{|t|^{2}}{2\pi
\alpha}\lambda^{\frac{3}{\alpha}-2\gamma}\frac{\Gamma\left(
\frac{3}{2\alpha}\right)\Gamma\left(\gamma-\frac{3}{2\alpha}\right)}{\Gamma(\gamma)}+o(|t|^2).\nonumber
\end{align}This shows that after crossing the point
$\alpha\gamma=3/2$, the leading behavior of $\sigma^2(t)$ is of
order $|t|^2$, which does not  depend on the parameters $\alpha$ and
$\gamma$.

\noindent Case III. The borderline case $\alpha\gamma=3/2$ is more
complicated. First, we find as in Case I that
\begin{align*}
\sigma^2(t)=\frac{4|t|^{2}}{\pi}\int_0^{\infty}
\frac{\sin^2(\omega/2)}{(\omega^{2\alpha}+\lambda^2|t|^{2\alpha})^{\gamma}}d\omega.
\end{align*}The integral \begin{align}\label{eq3_23_4}\int_0^{\infty}
\frac{\sin^2(\omega/2)}{(\omega^{2\alpha}+\lambda^2|t|^{2\alpha})^{\gamma}}d\omega\end{align}
does not converge when $t\rightarrow 0$ because of the singularity
at the origin of the integrand
$\omega^{-2\alpha\gamma}\sin^2\left(\omega/2\right)$. Since $\sin
z\sim z$ when $z\rightarrow 0$, we write \eqref{eq3_23_4} as a sum
of two terms $A_1(t)$ and $A_2(t)$, where
\begin{align*}
A_1(t)=&\int_0^{1} \frac{
(\omega/2)^2}{(\omega^{2\alpha}+\lambda^2|t|^{2\alpha})^{\gamma}}d\omega,
\end{align*}reflects the divergence of \eqref{eq3_23_4} when $t\rightarrow 0$; and \begin{align*}A_2(t)=\int_0^{1}
\frac{\sin^2(\omega/2)-(\omega/2)^2}{(\omega^{2\alpha}+\lambda^2|t|^{2\alpha})^{\gamma}}d\omega
+\int_1^{\infty} \frac{
\sin^2(\omega/2)}{(\omega^{2\alpha}+\lambda^2|t|^{2\alpha})^{\gamma}}d\omega\end{align*}
carry the finite part.  By making a change of variable
$v=\omega^{2\alpha}$ or equivalently $\omega=v^{\gamma/3}$, we find
that
\begin{align*}
A_1(t)=\frac{1}{8\alpha}\int_0^1\frac{v^{\gamma-1}dv}{(v+\lambda^2|t|^{2\alpha})^{\gamma}}.
\end{align*}To find the asymptotic behavior of $A_1(t)$ as $t\rightarrow 0$, we
split it again
 as the sum of $A_3(t)$ and $A_4(t)$,
where
\begin{align}\label{eq12_21_4}
A_3(t)=\frac{1}{8\alpha}\int_0^1\frac{dv}{(v+\lambda^2|t|^{2\alpha})}\sim
\frac{1}{4}\log\frac{1}{|t|}-\frac{1}{4\alpha}\log\lambda+o(1)
\end{align}give the divergence part; and $A_4(t):=A_1(t)-A_3(t)$
gives
a finite limit:
\begin{align}\label{eq12_21_5}
A_4(t)=&\frac{1}{8\alpha}\int_0^1\frac{v^{\gamma-1}-(v+\lambda^2|t|^{2\alpha})^{\gamma-1}}
{(v+\lambda^2|t|^{2\alpha})^{\gamma}}dv=\frac{1}{8\alpha}\int_0^{\frac{1}{\lambda^2|t|^{2\alpha}}}
\frac{v^{\gamma-1}-(1+v)^{\gamma-1}}{(1+v)^{\gamma}}dv\\
\sim &\frac{1}{8\alpha}\int_0^{\infty}
\frac{v^{\gamma-1}-(1+v)^{\gamma-1}}{(1+v)^{\gamma}}dv+o(1)=
\frac{1}{8\alpha}\left(\psi(1)-\psi(\gamma)\right)+o(1)\nonumber.
\end{align}In the last equality, we have used \#3.219 of \cite{15}, with
 $\psi(z)=\Gamma'(z)/\Gamma(z)$ being the logarithmic derivative of
the gamma function. The limit of  $A_2(t)$ when $t\rightarrow 0$ is
given by
\begin{align*}
A_2(0)=\int_0^1\omega^{-3}\left(\sin^2\left(\frac{\omega}{2}\right)
-\left(\frac{\omega}{2}\right)^2\right)d\omega+\int_1^{\infty}
\omega^{-3}\sin^2\left(\frac{\omega}{2}\right)d\omega.
\end{align*}It can be computed explicitly by regularization method:
\begin{align}\label{eq12_21_6}
A_2(0)=&\lim_{\vep\rightarrow
0^+}\left\{\int_0^{\infty}\omega^{-3+\vep}\sin^2\left(\frac{\omega}{2}\right)
d\omega
-\frac{1}{4}\int_0^1 \omega^{-1+\vep}d\omega\right\}\\
=&\lim_{\vep\rightarrow
0^+}\left\{\frac{1}{2}\cos\frac{\pi\vep}{2}\frac{\Gamma(1+\vep)}{\vep(1-\vep)(2-\vep)}-\frac{1}{4\vep}\right\}=
\frac{1}{4}\left(\psi(1) +\frac{3}{2}\right).\nonumber
\end{align}
Combining \eqref{eq12_21_4}, \eqref{eq12_21_5} and
\eqref{eq12_21_6}, we find that the leading behavior of
$\sigma^2(t)$ when $\alpha\gamma=3/2$ is:
\begin{align}\label{eq12_24_6}
\sigma^2(t)\sim\frac{|t|^2}{\pi}\log\frac{1}{|t|}
-\frac{|t|^2}{\pi}\left\{\frac{1}{\alpha}\log\lambda+\frac{1}{2\alpha}\left(\psi(\gamma)-\psi(1)\right)
-\psi(1)-\frac{3}{2}\right\}+o(1).
\end{align}This shows that the leading behavior of $\sigma^2(t)$ at the borderline case $\alpha\gamma=3/2$ is of order
$|t|^2\log(1/|t|)$.

\begin{figure}\centering \epsfxsize=.52\linewidth
\epsffile{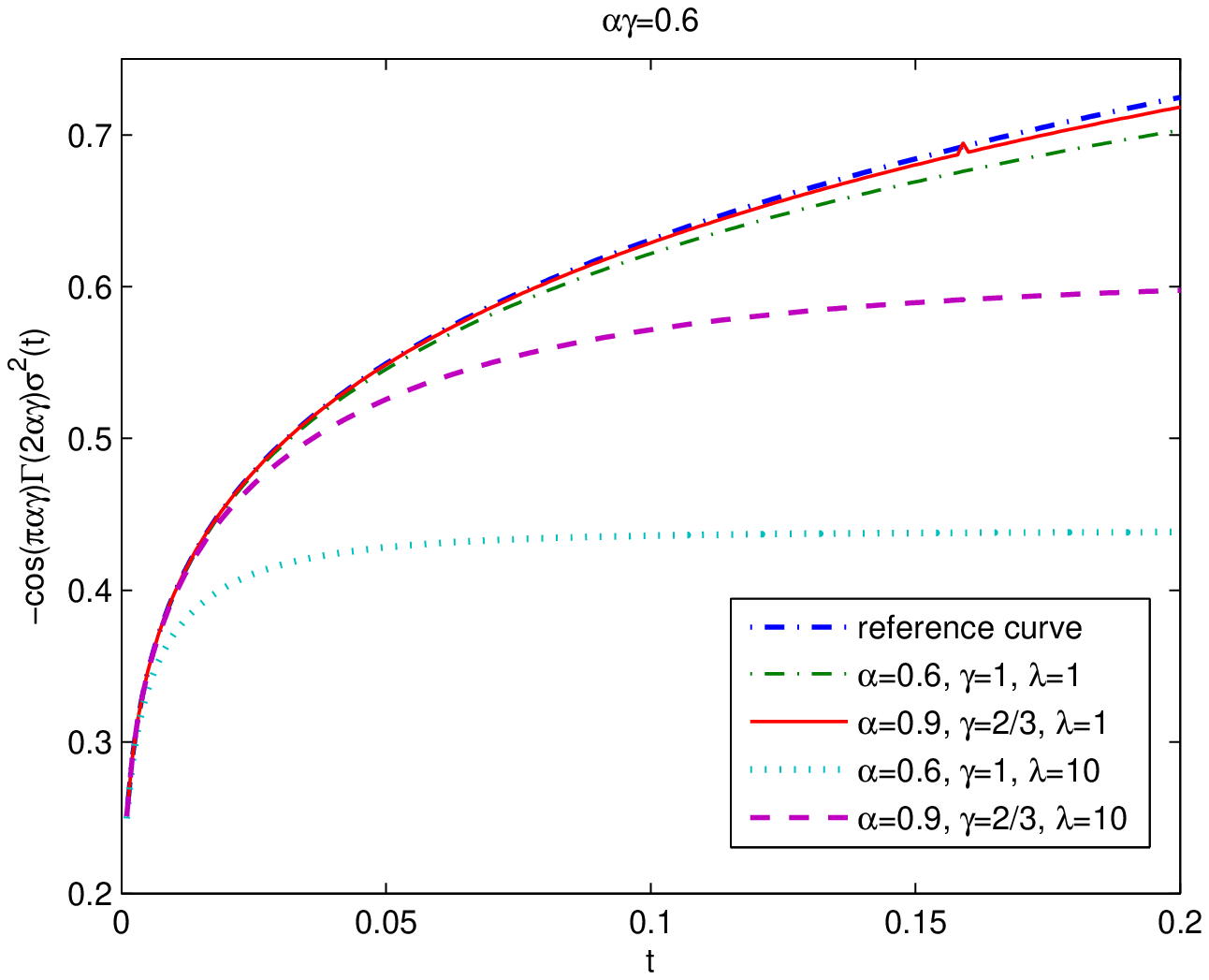}\centering \epsfxsize=.52\linewidth
\epsffile{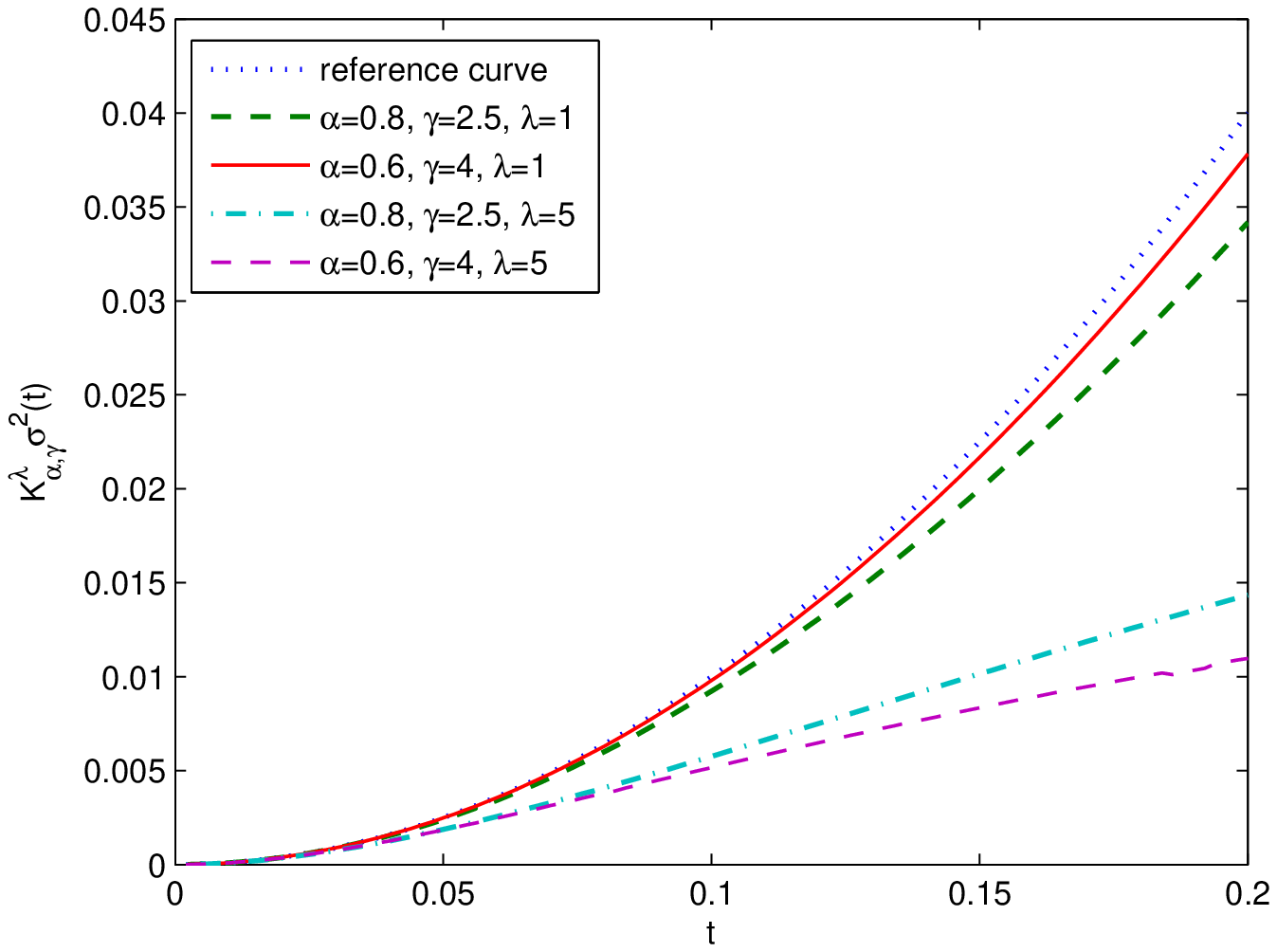}\caption{These graphs show the small time
behavior of $\sigma^2(t)$. Left: The function
$-\cos(\pi\alpha\gamma)\Gamma(2\alpha\gamma)\sigma^2(t)$ is plotted
as a function of $t$. The reference curve is
$y=|t|^{2\alpha\gamma-1}$. Here $\alpha\gamma=0.6$. The graph shows
that $-\cos(\pi\alpha\gamma)\Gamma(2\alpha\gamma)\sigma^2(t)\sim
|t|^{2\alpha\gamma-1}$ when $t\rightarrow 0$. Right: The function
$K_{\alpha,\gamma}^{\lambda}\sigma^2(t)$ is plotted as a function of
$t$, where
$K_{\alpha,\gamma}^{\lambda}=2\pi\alpha\lambda^{2\gamma-\frac{3}{\alpha}}
\frac{\Gamma(\gamma)}{\Gamma\left(\frac{3}{2\alpha}\right)
\Gamma\left(\gamma-\frac{3}{2\alpha}\right)}$. The reference curve
is $y=t^2$. The graph shows that
$K_{\alpha,\gamma}^{\lambda}\sigma^2(t)\sim t^2$ when $t\rightarrow
0$.}\end{figure}

\begin{figure}\centering \epsfxsize=.6\linewidth
\epsffile{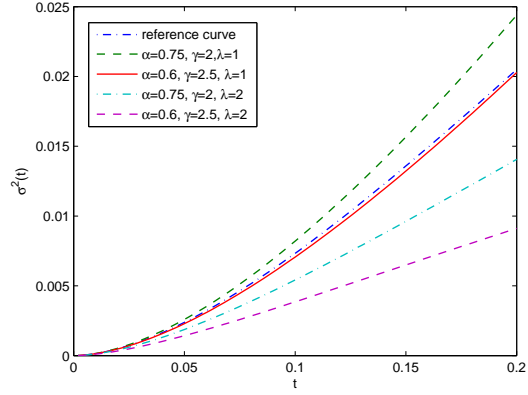}\caption{The graph shows the small time behavior
of $\sigma^2(t)$ when $\alpha\gamma=3/2$. The reference curve is
$t^2\log(1/|t|)$.}\end{figure}

    The behavior of $\sigma^2(t)$ at small $t$ is
illustrated graphically in Figure 2 and Figure 3.     We can further
confirm our results by checking the small time behavior of
$\sigma^2(t)$
   for the case $\alpha=1$, with the
 covariance function $C_{1,\gamma}(t)$   given explicitly by \eqref{eq8}.
 By using \#8.445, 8.446 and 8.485 of Ref. \cite{15} one gets
for $\nu\notin\Z$, \begin{align}\label{eq12_21_7}
K_{\nu}(z)=K_{-\nu}(z)=\frac{\pi}{2\sin (\pi
\nu)}\left\{\sum_{j=0}^{\infty}
\frac{(z/2)^{2j-\nu}}{j!\Gamma(j+1-\nu)}-\sum_{j=0}^{\infty}\frac{(z/2)^{2j+\nu}}{j!
\Gamma(j+1+\nu)}\right\};
\end{align}and in the case $\nu=\pm m$, $m$ a nonnegative integer,
\begin{align}\label{eq12_21_8}
K_{\nu}(z)=&\frac{1}{2}\sum_{j=0}^{m-1}\frac{(-1)^j
(m-j-1)!}{j!}\left(\frac{z}{2}\right)^{2j-m}
\\&+(-1)^{m+1}\sum_{j=0}^{\infty}\frac{(z/2)^{m+2j}}
{j!(m+j)!}\left\{\log\frac{z}{2}-\frac{1}{2}\psi(j+1)-\frac{1}{2}\psi(j+1+m)\right\}.\nonumber
\end{align}
  From \eqref{eq12_21_7} and \eqref{eq12_21_8}, one finds that the
  variance of $X_{1,\gamma}(t)$ is
$$C_{1,\gamma}(0)=\frac{\sqrt{\pi}}{2\lambda^{2\gamma-1}\sin\left(\pi(\gamma-\frac{1}{2}\right)\Gamma(\gamma)
\Gamma\left(\frac{3}{2}-\gamma\right)};$$ and
  the  leading  behavior of
$\sigma^2(t)$ as $t\rightarrow 0$ is given by:

\noindent $\bullet$\;\;  If $1/2<\gamma<3/2$,
\begin{align}\label{eq12_24_1}
\sigma^2(t)\sim & \frac{\sqrt{\pi}|t|^{2\gamma-1}}{2^{2\gamma-1}
\sin\left(\pi\left(\gamma-\frac{1}{2}\right)\right)\Gamma(\gamma)\Gamma\left(\gamma-\frac{1}{2}\right)}\\
=&-\frac{|t|^{2\gamma-1}}{\cos(\pi\gamma)\Gamma(2\gamma)}.\nonumber
\end{align}In the last step, we have used the identity $\Gamma(2z)=(2^{2z-1}/\sqrt{\pi})\Gamma(z)\Gamma(z+(1/2))$.

\noindent $\bullet$\;\;   If $\gamma>3/2$,
\begin{align}\label{eq12_24_2}
\sigma^2(t)\sim
\frac{1}{4\sqrt{\pi}}\frac{\Gamma\left(\gamma-\frac{3}{2}\right)}{\Gamma(\gamma)}|t|^2.
\end{align}

\noindent $\bullet$\;\;   If $\gamma=3/2$,
\begin{align}\label{eq12_24_3}
\sigma^2(t)\sim
\frac{|t|^2}{\pi}\left(\log\frac{1}{|t|}-\log\lambda+\log
2+\psi(1)+\frac{1}{2}\right).
\end{align}

 By putting $\alpha=1$ in the small time asymptotic formulas
\eqref{eq12_24_5}, \eqref{eq12_24_4} and \eqref{eq12_24_6} of
$\sigma^2(t)$, one obtains the formulas \eqref{eq12_24_1} ---
\eqref{eq12_24_3}, which confirm our results.

Next, we study the asymptotic behavior of   $C_{\alpha,\gamma}(t)$
for $t\rightarrow \infty$. When $0<\alpha<1$, we can make use of
\eqref{eq12_21_2}. Making a change of variable $u\mapsto u/t$ and
using the formula
$$\frac{1}{(1+z)^{\gamma}}=\sum_{j=0}^{\infty}\frac{(-1)^j
\Gamma(\gamma+j)}{j!\Gamma(\gamma)}z^j,$$ we can derive the
following $t\rightarrow \infty$ asymptotic expression for
$C_{\alpha,\gamma}(t)$ which is valid when $\alpha\in (0,1)$:
\begin{align}\label{eq12_20_3} C_{\alpha,\gamma}(t)=&
\frac{1}{\pi}\text{Im}\left\{t^{-1}\int_0^{\infty}\frac{e^{-u}du}{\left(\lambda^2+e^{-i\pi\alpha}\frac{u^{2\alpha}}
{t^{2\alpha}} \right)^{\gamma}}\right\}\\
\sim&\frac{1}{\pi}\text{Im}\left\{t^{-1}\int_0^{\infty}e^{-u}\sum_{j=0}^{\infty}
\frac{(-1)^j\Gamma(\gamma+j)}{j!\Gamma(\gamma)}e^{-i\pi\alpha
j}\frac{u^{2\alpha j}} {t^{2\alpha j}}\lambda^{-2\gamma-2j}
\right\}\nonumber\\\nonumber \sim &\frac{1}{\pi\Gamma(\gamma)}
\sum_{j=1}^{\infty}\frac{(-1)^{j+1}\lambda^{-2(\gamma+j)}}{j!}
\Gamma(\gamma+j)\Gamma(1+2\alpha j)\sin\left(\pi\alpha
j\right)t^{-(2\alpha j+1)}.\end{align}
 The leading term is
\begin{align}\label{eq12_31_6}C_{\alpha,\gamma}(t)\sim \frac{\gamma}{\pi}\lambda^{-2(\gamma+1)}
\Gamma(1+2\alpha)\sin\left(\pi\alpha\right)t^{-(2\alpha+1)}.\end{align}
By letting $\lambda=1$, \eqref{eq12_20_3} is in agreement with the
analogous result given in Ref. \cite{14} for Linnik distribution.
Notice that when $0<\alpha<1$, $C_{\alpha,\gamma}(t)$ is of
polynomial decay with order $t^{-2\alpha-1}$ when $t\rightarrow
\infty$ (see Figures 4 and 5).

\begin{figure}\centering \epsfxsize=.6\linewidth
\epsffile{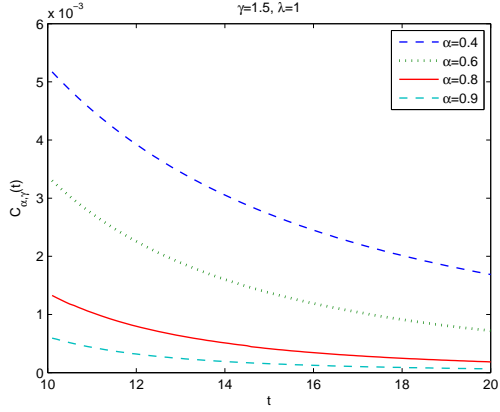} \caption{This graph shows the large time
behavior of $C_{\alpha,\gamma}(t)$.}
\end{figure}
\begin{figure}\centering \epsfxsize=.52\linewidth
\epsffile{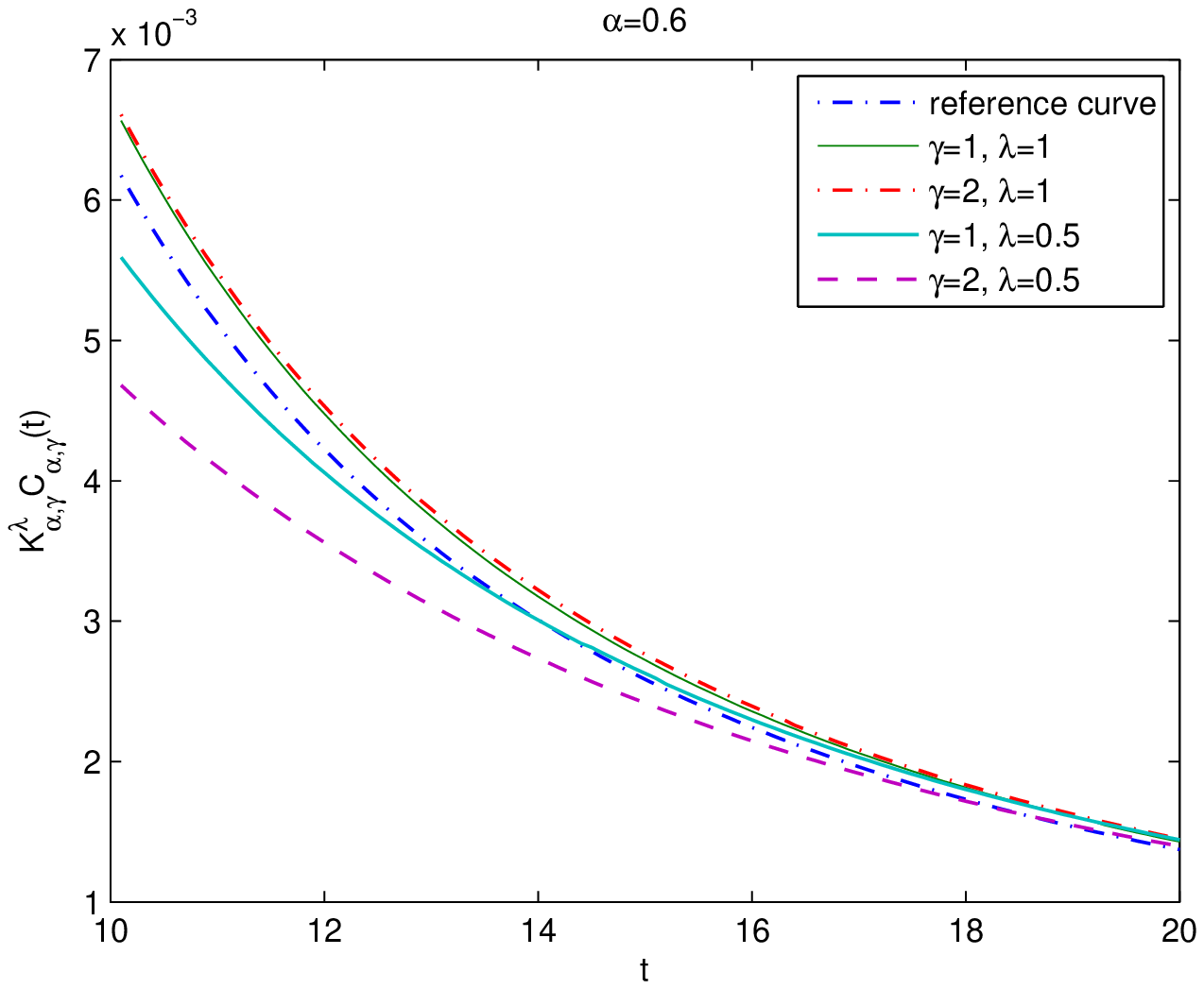}\centering \epsfxsize=.52\linewidth
\epsffile{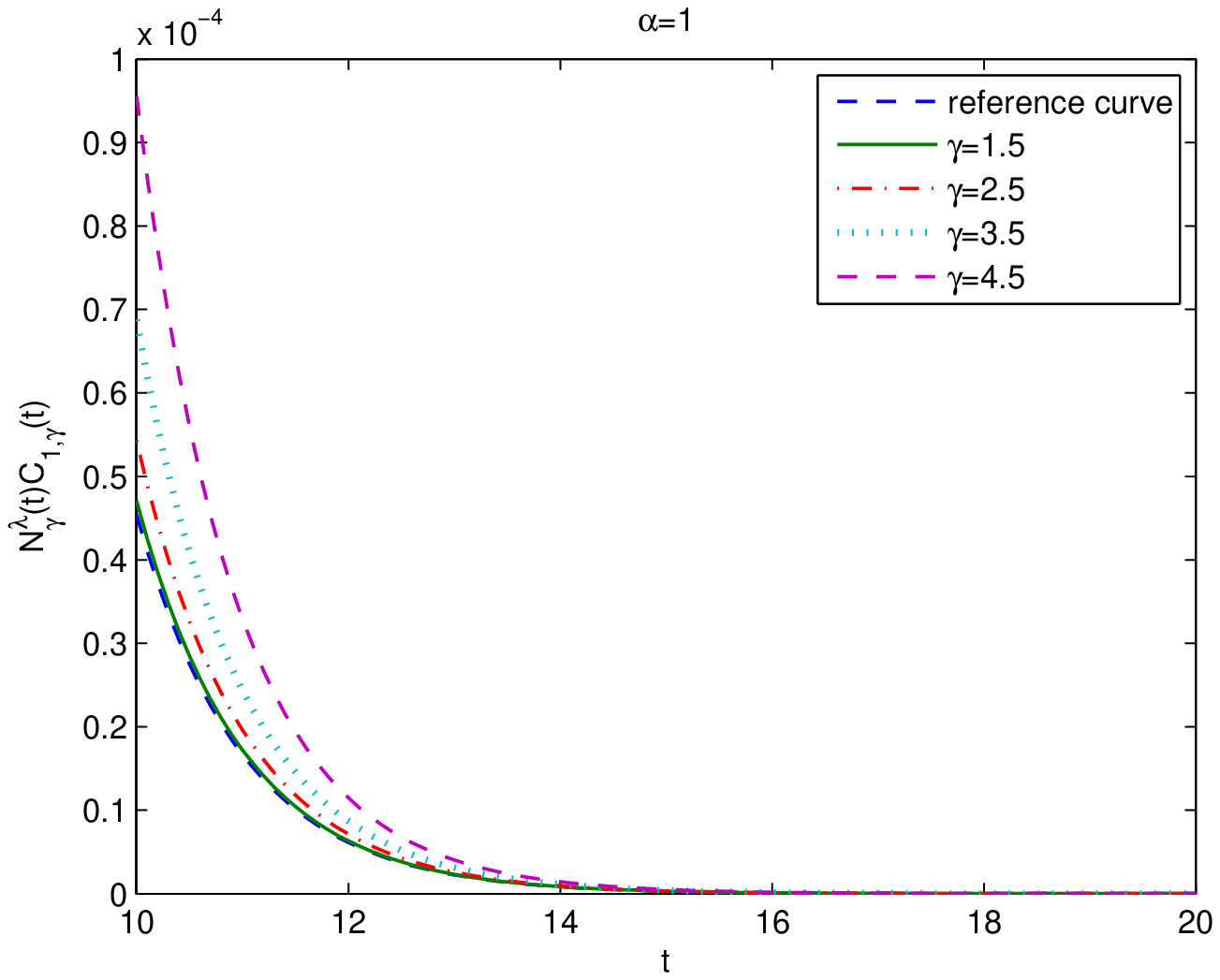}\caption{The large time behavior of $C(t)$ when
$\alpha=0.6$ and $\alpha=1$ respectively. Left: Here
$K_{\alpha,\gamma}^{\lambda}=\frac{\lambda^{2(\gamma+1)}\pi}{\gamma\Gamma(1+2\alpha)\sin(\pi\alpha)}$
 and the reference curve is $y=t^{-(2\alpha+1)}$. The graph shows that $K_{\alpha,\gamma}^{\lambda}C_{\alpha,\gamma}
 (t)\sim t^{-2\alpha-1}$ as $t\rightarrow \infty$. Right: Here $N_{\gamma}^{\lambda}(t)
 =(2\lambda)^{\gamma}\Gamma(\gamma)t^{\gamma-1}$, $\lambda=1$ and the reference curve is $y=e^{-\lambda t}$.
 The graph shows that $N_{\gamma}^{\lambda}(t)C_{1,\gamma}(t)\sim e^{-\lambda t}$ for large $t$.}\end{figure}

 The case $\alpha=1$ has to be considered separately.
Using the relation (\#8.451, no. 6, ref. \cite{15}):
\begin{align*}
K_{\nu}(z)=\sqrt{\frac{\pi}{2z}}e^{-z}\left(1+\frac{4\nu^2-1}{8z}+\ldots\right),
\end{align*} we find from the explicit formula \eqref{eq8} for
$C_{1,\gamma}(t)$ that
\begin{align}\label{eq12_31_8}
C_{1,\gamma}(t)\sim
\frac{|t|^{\gamma-1}}{(2\lambda)^{\gamma}\Gamma(\gamma)}e^{-\lambda|t|}\hspace{1cm}\text{as}\;\;
|t|\rightarrow \infty.
\end{align}One notice that at large time, $C_{1,\gamma}(t)$ decays
exponentially, in contrast the large time behavior of
$C_{\alpha,\gamma}(t)$, $\alpha\in (0,1)$ \eqref{eq12_20_3}, which
decays polynomially (see Figure 5).

From the above results, it appears that the small time asymptotic
behavior of the covariance $C_{\alpha,\gamma}(t)$ varies as
$|t|^{\min\{2\alpha\gamma-1,2\}}$, depending on both $\alpha$ and
$\gamma$. However, if the index $\gamma$ is replaced by
$\gamma/\alpha$, then $C_{\alpha,\gamma}(t)\sim
|t|^{\min\{2\gamma-1, 2\}}$ as $t\rightarrow 0$. Thus, together with
the large time asymptotic behavior
$C_{\alpha,\gamma}(t)\sim|t|^{-2\alpha-1}$ as $t\rightarrow \infty$,
we have the result that the small and large time asymptotic behavior
of the covariance of $X_{\alpha,\gamma}(t)$ are separately
characterized by $\gamma$ and $\alpha$. The physical implications of
this result will be discussed in the subsequent sections.

\subsection{Locally Asymptotically Self-Similarity and Fractal Dimension}

Recall that a stationary random process cannot be self-similar
\cite{rr1}. It would be interesting to see whether
$X_{\alpha,\gamma}(t)$ satisfies a weaker self-similar property,
namely self-similarity at very small time scales. First we introduce
some definitions. A positive function $f$ is asymptotically
homogeneous of order $\kappa$ at $\infty$   if there exists a
non-zero function $f^{\infty}$ such that, for almost every
$\omega\in\R$ and $r>0$, $f^r(\omega)=r^{-\kappa}f(r\omega)$ has a
limit $f^{\infty}(\omega)$ when $r\rightarrow\infty$. Clearly,
$f^{\infty}(\omega)$ is homogeneous of order $\kappa$, thus fixes
the index $\kappa$ uniquely. One can easily verify that the spectral
density $S_{\alpha,\gamma}(\omega)$ is asymptotically homogeneous of
order $2\alpha\gamma$ at  $\infty$, with
$S_{\alpha,\gamma}^{\infty}(\omega)=\omega^{-2\alpha\gamma}/(2\pi)$.
In addition, the spectral density satisfies the following property:
there exist positive constants $A, B\in\R$ such that
$S_{\alpha,\gamma}(\omega)\leq B|\omega|^{-2\alpha\gamma}$, for
almost all $|\omega|>A$. This is clearly true since $\lambda>0$
implies
$$S_{\alpha,\gamma}(\omega)=\frac{1}{2\pi}\frac{1}{(\lambda^2+\omega^{2\alpha})^{\gamma}}
<\frac{|\omega|^{-2\alpha\gamma}}{2\pi}.$$  Using a result
(Proposition 2 in Ref. \cite{rr2}), one concludes that if
$\alpha\gamma\in (1/2,3/2)$, the fractional process
$X_{\alpha,\gamma}(t)$ is locally asymptotically self-similar (LASS)
of order $\alpha\gamma-1/2$, that is for $u\in\R$,
\begin{align*}\lim_{\vep\rightarrow 0^+}\left\{\frac{X_{\alpha,\gamma}(t_0+\vep u)-X_{\alpha,\gamma}(t_0)}
{\vep^{\alpha\gamma-\frac{1}{2}}}\right\}=\Bigl\{X_{\alpha,\gamma}^{\infty}(u)\Bigr\},\end{align*}
with the convergence in the sense of distribution on the space of
continuous paths on $\R$. Note that this result is in agreement with
\eqref{eq12_24_5} which asserts that the covariance $\sigma^2(t)\sim
C|t|^{2\alpha\gamma-1}$ as $t\rightarrow 0$. The limit process or
the tangent process $X_{\alpha,\gamma}^{\infty}$ is self-similar of
order $\alpha\gamma-\frac{1}{2}$. It can be identified with
fractional Brownian motion if $\alpha\gamma=H+1/2$, where $H$
denotes the Hurst index of the fractional Brownian motion. Just like
the case of ordinary oscillator process, which locally behaves like
Brownian motion, likewise the fractional oscillator process
$X_{\alpha,\gamma}(t)$ has the same local behavior as fractional
Brownian motion of order $\alpha\gamma-1/2$. In fact,
\eqref{eq12_24_5} gives us
\begin{align*}
&\left\langle \lim_{\vep\rightarrow
0^+}\left[\frac{X_{\alpha,\gamma}(t_0+\vep
u)-X_{\alpha,\gamma}(t_0)} {\vep^{\alpha\gamma-\frac{1}{2}}}\right]
\left[\frac{X_{\alpha,\gamma}(t_0+\vep v)-X_{\alpha,\gamma}(t_0)}
{\vep^{\alpha\gamma-\frac{1}{2}}}\right]\right\rangle\\
=&\lim_{\vep\rightarrow 0} \frac{\sigma^2(\vep u)+\sigma^2(\vep v)
-\sigma^2(\vep(u-v))}{2\vep^{2\alpha\gamma-1}}\\
=&-\frac{1}{2\cos(\pi\alpha\gamma)\Gamma(2\alpha\gamma)}\left(|u|^{2\alpha\gamma-1}+|v|^{2\alpha\gamma-1}
-|u-v|^{2\alpha\gamma-1}\right),
\end{align*}which is the covariance function $\langle B_H(u)B_H(v)\rangle$ for the fractional
Brownian motion
$$B_H(u):=\frac{1}{\sqrt{2\pi}}\int\limits_{\R}\frac{e^{i\omega
t}-1}{|\omega|^{H+\frac{1}{2}}}d\omega,$$ if we identify $H$ with
$\alpha\gamma-1/2$ (see Figure 6). \begin{figure}
\centering\epsfxsize=.52\linewidth \epsffile{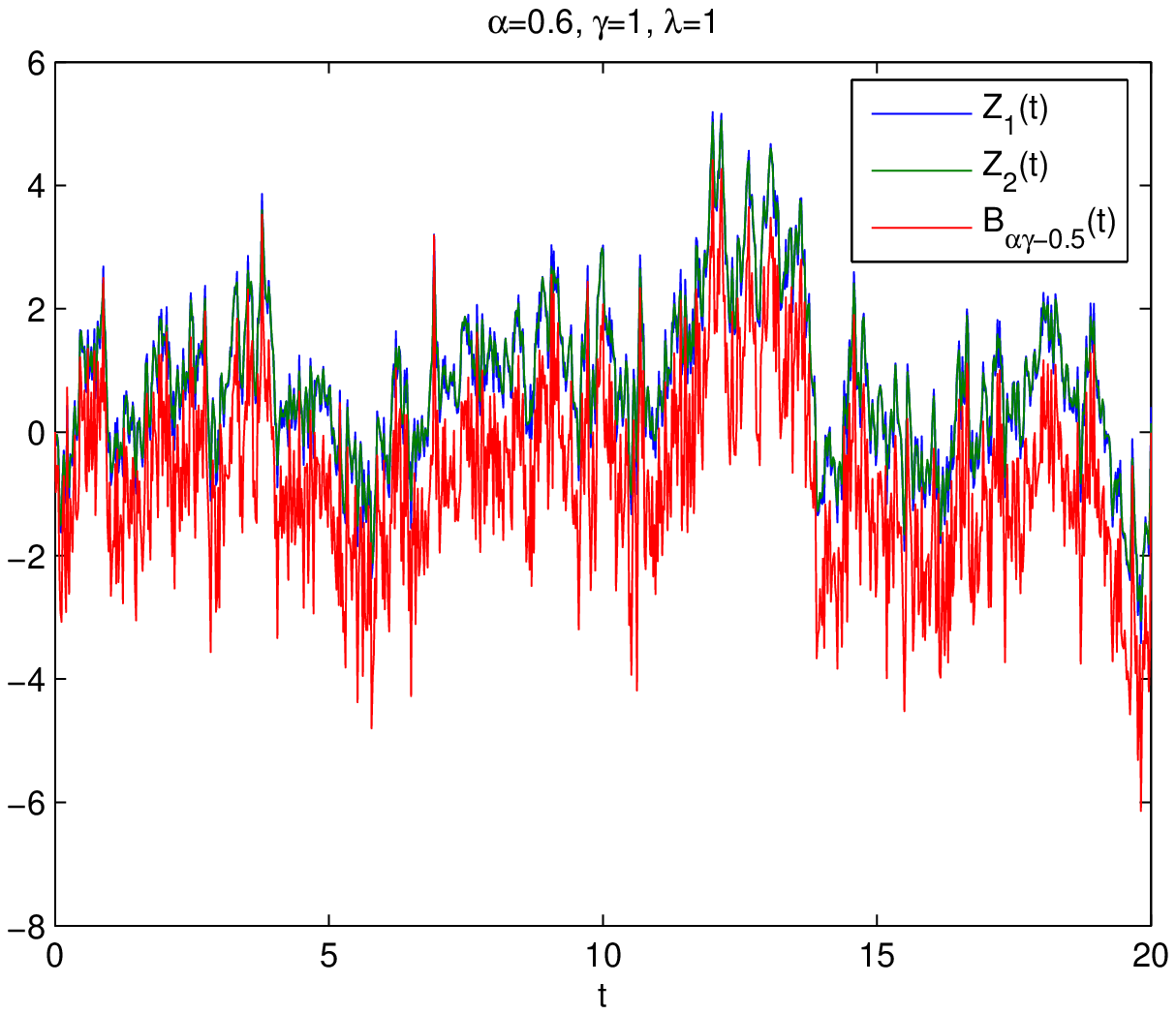}\centering
\epsfxsize=.52\linewidth \epsffile{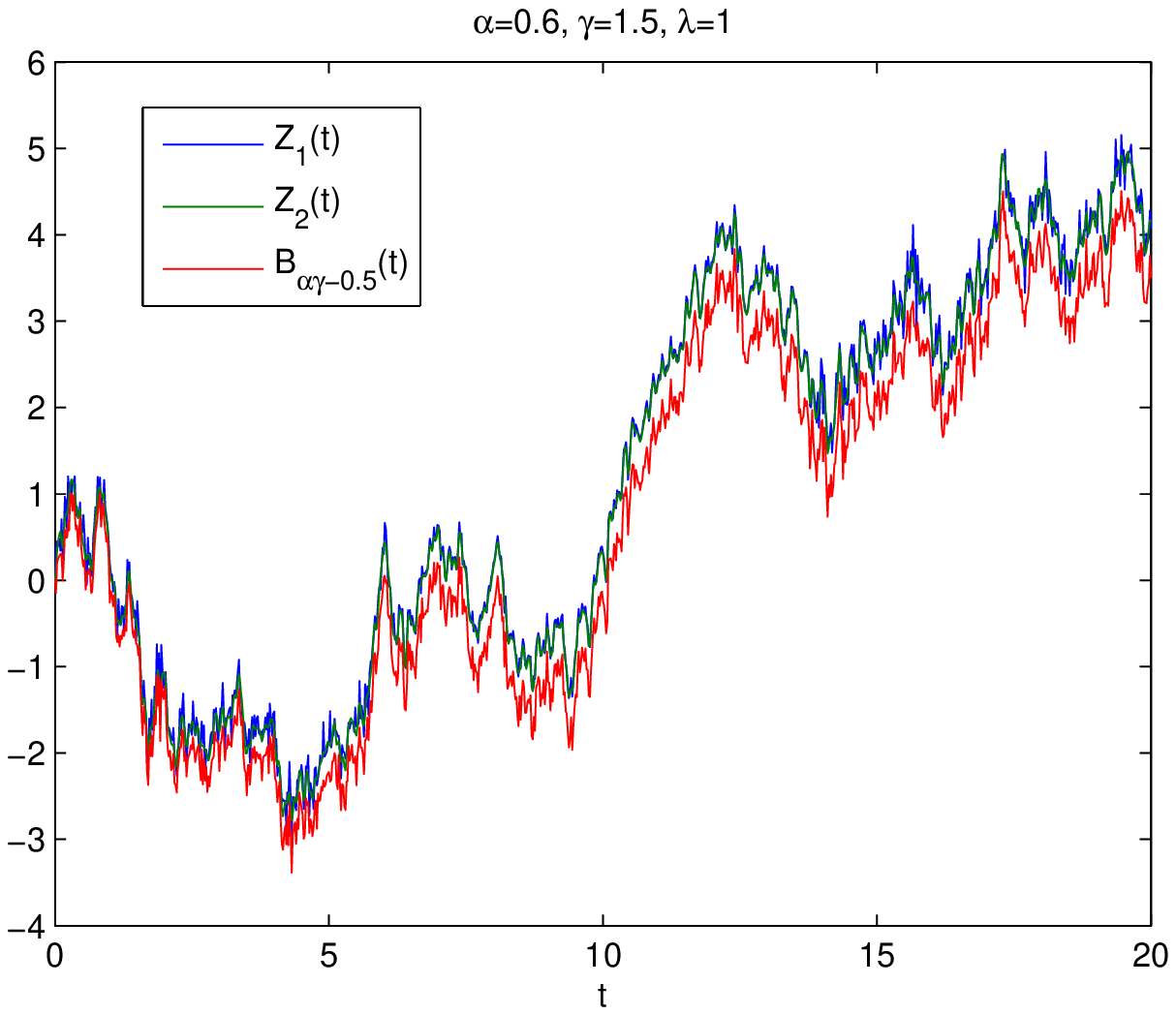}\caption{The graphs
shows that
$\vep^{-\alpha\gamma+\frac{1}{2}}\left[X_{\alpha,\gamma}(\vep t)
-X_{\alpha,\gamma}(0)\right]$ approaches the fractional Brownian
motion $B_{\alpha\gamma-\frac{1}{2}}(t)$ when $\vep\rightarrow 0$.
Here
$Z_i(t)=\vep_i^{-\alpha\gamma+\frac{1}{2}}\left[X_{\alpha,\gamma}(\vep_i
t) -X_{\alpha,\gamma}(0)\right]$ with $\vep_1=0.0001$,
$\vep_2=0.00005$, $i=1,2$.}
\end{figure} We also remark that when
$\alpha\beta=3/2$, we find from \eqref{eq12_24_6} that the
fractional oscillator process $X_{\alpha,\gamma}(t)$ fails to
satisfy the LASS property. When $\alpha\gamma$ exceeds $3/2$, the
process $X_{\alpha,\beta}(t)$ becomes differentiable with variance
\begin{align*}
&\lim_{\vep\rightarrow 0}\left\langle
\left[\frac{X_{\alpha,\gamma}(t_0+\vep)-X_{\alpha,\gamma}(t_0)}{\vep}\right]^2\right\rangle
=\lim_{\vep\rightarrow
0}\frac{\sigma^2(\vep)}{\vep^2}=\frac{\lambda^{\frac{3}{\alpha}-2\gamma}}{2\pi
\alpha}\frac{\Gamma\left(
\frac{3}{2\alpha}\right)\Gamma\left(\gamma-\frac{3}{2\alpha}\right)}{\Gamma(\gamma)},
\end{align*}which follows from \eqref{eq12_24_4}.

Another important concept in the study of the sample path properties
of a stochastic process  is the H\"olderian property. A function
$f:[a,b]\rightarrow \R$ is H\"olderian of order $\kappa\in (0,1]$ if
$$|f(t)-f(s)|\leq K |t-s|^{\kappa}, \hspace{1cm}\text{for all}\; s,t\in
[a,b]$$for some constant $K>0$. It is well-known that if $Z(t)$ is a
stationary process and $\sigma^2(t)=\langle
\left[Z(t)-Z(0)\right]^2\rangle$ satisfies $$\sigma^2(t)\leq
C|t|^{2\kappa},$$ then almost surely (a.s.) the sample path of
$Z(t)$ is H\"olderian of order $\kappa-\vep$ for all $\vep>0$
\cite{30, AL, ALV}. Applying this concept to the fractional
oscillator process $X_{\alpha,\gamma}(t)$, we find from
\eqref{eq12_24_5}, \eqref{eq12_24_4} and \eqref{eq12_24_6} that for
any $\vep>0$, the sample path of $X_{\alpha,\gamma}(t)$ is
H\"olderian of order $\alpha\gamma-1/2-\vep$ if $\alpha\gamma<3/2$,
and is H\"olderian of order $1-\vep$ if $\alpha\gamma\geq 3/2$.

 Now we want to
consider the fractal dimension $D$ of the graph for the fractional
oscillator process $X_{\alpha,\gamma}(t)$. Since fractal dimension
is a local concept, fractality is defined for infinitesimally small
time scales. For a locally self-similar process, one may apply the
following result to obtain its fractal dimension. A process which is
LASS of order $\kappa>0$ and its sample paths are a.s.  $\kappa
-\vep$--H\"olderian  for all $\vep>0$, then the fractal dimension of
its graph is a.s. equals to $2-\kappa$ \cite{30, 21}. Applying this
result to $X_{\alpha,\gamma}(t)$ gives the fractal dimension
$D=\frac{5}{2}-\alpha\gamma$ for the graph of the fractional
oscillator process when $1/2<\alpha\gamma<3/2$. For
$\alpha\gamma\geq 3/2$,  the fractal dimension of the graph of
$X_{\alpha,\gamma}(t)$ is equal to $1$. In other words, the fractal
dimension of the graph of $X_{\alpha,\gamma}(t)$ is $\max\left\{ 1,
\frac{5}{2}-\alpha\gamma\right\}$. Again, if we replace $\gamma$ by
$\gamma/\alpha$, then the fractal dimension becomes $\max\left\{ 1,
\frac{5}{2}-\gamma\right\}$, which depends solely on $\gamma$.

\subsection{Short Range Dependence Property} First we recall that
the Ornstein--Uhlenbeck process or ordinary oscillator process (up
to a multiplicative constant) is the only stationary Gaussian Markov
process, thus rule out the possibility of the fractional oscillator
processes $X_{\alpha,\gamma}(t)$ being Markovian. Now we want to
find out the nature of memory possessed by $X_{\alpha,\gamma}(t)$,
whether it has long memory or long range-dependence (LRD), or short
memory or short-range dependence (SRD). A stationary Gaussian
process with covariance $C(t)$  is said to be LRD if for some finite
$t_{*}\geq 0$,
\begin{align}\label{eq28}
\int_{t_{*}}^{\infty}|C(t)|d\tau=\infty,
\end{align} otherwise it is SRD. From the results
\eqref{eq12_31_6} and \eqref{eq12_31_8},  we find that the
covariance of $X_{\alpha,\gamma}(t)$ behaves asymptotically as
$C_{\alpha,\gamma}(t)\sim t^{-(2\alpha+1)}$ if $\alpha\in (0,1)$ and
as $C_{\alpha,\gamma}(t)\sim e^{-\lambda t}t^{\gamma-1}$ if
$\alpha=1$, for $t\rightarrow \infty$. This shows that the
corresponding integral \eqref{eq28} is convergent and therefore the
fractional oscillator process $X_{\alpha,\gamma}(t)$ has SRD. It is
interesting to note that when $\alpha\in (0,1)$ the asymptotic order
of the covariance does not depend on the parameter $\gamma$.
Therefore, the parameter $\alpha$ characterize the asymptotic order
of the covariance $C_{\alpha,\gamma}(t)$ as $t\rightarrow \infty$.
Combining with the remark given earlier, one notes that it is
possible to separately characterize the fractal dimension and short
range dependence of $X_{\alpha,\gamma}(t)$ with two different
indices.

\section{Fluctuation--Dissipation Relation}

One of the important theorems in statistical mechanics is the
fluctuation-dissipation theorem \cite{New4} which relates the
coefficient of the covariance of the external random force in the
Langevin equation with the frictional coefficient. It would be
interesting to see whether there exists some kind of
fluctuation-dissipation relation for the fractional process
$X_{\alpha,\gamma}(t)$. For this purpose we re-express the
fractional Langevin equation \eqref{eq3_23_1} as
\begin{align}\label{eq3_23_1_2}
\left(\mathbf{D}_t^{2\alpha}+\lambda^{2\alpha}\right)^{\frac{\gamma}{2}}X_{\alpha,\gamma}(t)=\eta(t),
\end{align}
with the covariance of white noise $\eta(t)$ given by $\langle
\eta(t)\eta(s)\rangle =2B\delta(t-s)$, where $B$ is a constant
coefficient. If we regard $X_{\alpha,\gamma}(t)$ as the velocity
process, then the assumption of the thermalization of the fractional
velocity process based on the fractional generalization of the
classical principle of equipartition of energy \cite{New4} gives
\begin{align}\label{eq4_16_2} \left\langle
\left[X_{\alpha,\gamma}(t)\right]^2\right\rangle =
(kT)^{\alpha\gamma},
\end{align} where we have assumed the particle under consideration has unit
mass; $k$ is the Boltzmann constant and $T$ denotes temperature.
Under the condition $\alpha\gamma>1/2$, the process
$X_{\alpha,\gamma}(t)$ has finite variance given by
\eqref{eq4_16_1}:
\begin{align}\label{eq4_18_1}
C_{\alpha,\gamma}(0) =& \frac{B}{\pi
\alpha}\frac{\Gamma\left(\frac{1}{2\alpha}\right)
\Gamma\left(\gamma-\frac{1}{2\alpha}\right)}{\Gamma(\gamma)}\lambda^{1-2\alpha\gamma}.\end{align}
 From \eqref{eq4_16_2} and \eqref{eq4_18_1} one obtains
\begin{align}\label{eq4_16_3}
B=\frac{\pi\alpha\Gamma(\gamma)}{\Gamma\left(\frac{1}{2\alpha}\right)\Gamma\left(
\gamma-\frac{1}{2\alpha}\right)}\lambda^{2\alpha\gamma-1}(kT)^{\alpha\gamma}
=n(\alpha,\gamma)\lambda^{2\alpha\gamma-1}(kT)^{\alpha\gamma},
\end{align}
with $$n(\alpha,\gamma)=
\frac{\pi\alpha\Gamma(\gamma)}{\Gamma\left(\frac{1}{2\alpha}\right)\Gamma\left(
\gamma-\frac{1}{2\alpha}\right)}.$$ \eqref{eq4_16_3} can be regarded
as the generalized fluctuation--dissipation relation for the
fractional velocity process. When $\alpha=1$   and $\gamma=1$,
\eqref{eq4_16_3} reduces to the fluctuation--dissipation relation
for the ordinary Ornstein-Uhlenbeck process $X_{1,1}(t)$:
\begin{align*}
B=\lambda kT.
\end{align*}

     We can show that for a short range dependent process such as
$X_{\alpha,\gamma}(t)$, the leading term for the large time behavior
of the variance of its mean--square displacement does not depend on
the covariance of $X_{\alpha,\gamma}(t)$. For illustration, let us
first consider the simple case with position process
$Y_{\alpha,\gamma}(t)$ linked to the velocity process by
\begin{align}\label{eq4_23_1}X_{\alpha,\gamma}(t)=\frac{dY_{\alpha,\gamma}(t)}{dt}.\end{align} If we
assume that $Y_{\alpha,\gamma}(0)=0$, then the  variance of the
position process $Y_{\alpha,\gamma}(t)$ is  given by
\begin{align}\label{eq4_16_5}
\left\langle \left[Y_{\alpha,\gamma}(t)\right]^2\right\rangle
=\int_0^t\int_0^t C_{\alpha,\gamma}(|s_1-s_2|)ds_2 ds_1.
\end{align}
By some calculus, we have
\begin{align}\label{eq4_17_8}
\left\langle \left[Y_{\alpha,\gamma}(t)\right]^2\right\rangle
=&\int_0^t\int_0^{s_1} C_{\alpha,\gamma}(s_1-s_2)ds_2
ds_1+\int_0^{t}\int_{s_1}^{t}
C_{\alpha,\gamma}(s_2-s_1)ds_1ds_2\\
=&\int_0^t\int_0^s C_{\alpha,\gamma}(\tau) d\tau ds
+\int_0^t\int_0^{s_2}C_{\alpha,\gamma}(s_2-s_1)ds_1 ds_2\nonumber\\
=& 2\int_0^t \int_0^s C_{\alpha,\gamma}(\tau) d\tau ds=2\int_0^t
(t-\tau)C_{\alpha,\gamma}(\tau)d\tau.\nonumber
              \end{align}
Since the integral $\int_0^{\infty}C_{\alpha,\gamma}(\tau)d\tau$ is
convergent, we find that in the long-time $t\gg 1$ limit,
\begin{align}\label{eq4_23_2} \left\langle
\left[Y_{\alpha,\gamma}(t)\right]^2\right\rangle \sim 2 \left[
\int_0^{\infty} C_{\alpha,\gamma}(\tau) d\tau\right] t
\end{align} which is just ordinary diffusion with diffusion constant
$$D=\int_0^{\infty} C_{\alpha,\gamma}(\tau)d\tau.$$
Since
\begin{align}\label{eq4_16_6}
2\int_0^{\infty} C_{\alpha,\gamma}(\tau)d\tau =2\pi
S_{\alpha,\gamma}(0)=2B\lambda^{-2\alpha\gamma},
\end{align}using \eqref{eq4_16_3},
with $(kT)^{\alpha\gamma}$ replaced by $kT$ in view of
\eqref{eq4_23_1}, one gets
\begin{align*}
D=\frac{\pi\alpha\Gamma(\gamma)}{\Gamma\left(\frac{1}{2\alpha}\right)\Gamma\left(
\gamma-\frac{1}{2\alpha}\right)}\frac{kT}{\lambda},\end{align*}
which reduces to the well-known Einstein relation
$$D=\frac{kT}{\lambda}$$for $\alpha=\gamma=1$. This simplified example shows that the long time behavior
of $C_{\alpha,\gamma}(\tau)$ does not show up in the leading term of
the long time asymptotic expression of the variance $\left\langle
\left[Y_{\alpha,\gamma}(t)\right]^2\right\rangle$. Its effect only
appears in the second leading term (see appendix). This is due to
the fact that $X_{\alpha,\gamma}(t)$ is a short range process with
its covariance $C(\tau)\sim \tau^{-\kappa}$, $\kappa>1$ for
$\tau\rightarrow \infty$. We remark that \eqref{eq4_23_2} is
consistent with the result obtained for the case with $\alpha>0,
\gamma=1$ if the usual velocity-displacement relation
\eqref{eq4_23_1} is used \cite{new2}.

     Now we consider the fractional case
with the velocity linked to the displacement by  the following
relation: \begin{align}
X_{\alpha,\gamma}(t)=\;_0D_t^{\chi}Y_{\alpha,\gamma}(t),\hspace{1cm}
\frac{1}{2}<\chi <\frac{3}{2},\end{align} where $_0D_t^{\chi}$
denotes Riemann-Liouville fractional derivative of order $\chi$. If
we further assume that
$\left._0D_{t}^{\chi-j}Y_{\alpha,\gamma}(t)\right|_{t=0}=0$ for
$j=1$ if $\chi\leq 1$ and $j=1,2$ if $\chi>1$, then the position
process is given by
\begin{align}\label{eq4_17_9}
Y_{\alpha,\gamma}(t)=\;_0I_t^{\chi}X_{\alpha,\gamma}(t)=\frac{1}{\Gamma(\chi)}\int_0^{t}
(t-u)^{\chi-1}X_{\alpha,\gamma}(u)du.
\end{align}
One can show that (see appendix) \begin{align}\label{eq4_16_7}
\left\langle
\left[Y_{\alpha,\gamma}(t)\right]^2\right\rangle=2B\lambda^{-2\alpha\gamma}\left[\frac{t^{2\chi-1}}{
(2\chi-1) \Gamma(\chi)^2}\right]+O\left(t^{\max\{0, 2\chi-2,
2\chi-2\alpha-1\}}\log t\right).\end{align} Note that the term in
the bracket on the right hand side of \eqref{eq4_16_7} is just the
variance of the Riemann-Liouville fractional Brownian motion indexed
by $\chi-1/2$ \cite{New7}. By setting $\chi=\alpha\gamma$ and
substituting $B$ from \eqref{eq4_16_3}, we get
\begin{align}\label{eq4_23_3} \left\langle
\left[Y_{\alpha,\gamma}(t)\right]^2\right\rangle\sim &
\frac{2\pi\alpha\Gamma(\gamma)}{\Gamma\left(\frac{1}{2\alpha}\right)\Gamma\left(
\gamma-\frac{1}{2\alpha}\right)}\frac{(kT)^{\alpha\gamma}}{\lambda}\left[\frac{t^{2\alpha\gamma-1}}{
(2\alpha\gamma-1) \Gamma(\alpha\gamma)^2}\right]\\ \sim &
N(\alpha,\gamma)\left(\frac{kT}{\lambda}\right)^{\alpha\gamma}
t^{2\alpha\gamma-1},\nonumber
\end{align} where  $N(\alpha,\gamma)$ is a constant term depending on $\alpha$
  and $\gamma$, which can be normalized to become unity. Let $P(x,t)$ denotes the probability
  distribution of the equilibrium displacement process (i.e. when $t\rightarrow \infty$),
then the effective Fokker-Planck equation is \begin{align*}
\frac{\pa }{\pa t} P(x,t)=D(t) \frac{\pa^2}{\pa x^2}P(x,t),
\end{align*} with the diffusion coefficient $D(t)$ given by
\cite{new2}
\begin{align*}
     D(t)=\frac{1}{2}\frac{\pa}{\pa t}\left\langle \left[Y_{\alpha,\gamma}(t)\right]^2\right\rangle.\end{align*}
For equilibrium state, $\left\langle
\left[Y_{\alpha,\gamma}(t)\right]^2\right\rangle$ is given by
\eqref{eq4_23_3}, so
\begin{align*}
D(t)=\frac{1}{2}\left(\frac{kT}{\lambda}\right)^{\alpha\gamma}(2\alpha\gamma-1)t^{2\alpha\gamma-2},
\end{align*}
which can be regarded as the effective diffusion coefficient for
fractional Brownian motion if we let $\alpha\gamma-1/2=H$, the Hurst
index \cite{New5, New5_2, New6}. Our result differs from that of
ref. \cite{New5, New5_2, New6} which has $(kT/\lambda)$ instead of
the $(kT/\lambda)^{\alpha\gamma}$ given above. This is due to our
use of $\lambda^{\alpha}$ in the Langevin equation
\eqref{eq3_23_1_2} and the fractional generalization of
equipartition principle \eqref{eq4_16_2}.

From the discussion above, we see that the long-time dependence of
the covariance of $X_{\alpha,\gamma}(t)$   does not enter in the
leading term of the variance of $Y_{\alpha,\gamma}(t)$. It only
appears as second leading term. Instead, the property of the leading
term depends on the differential relationship between
$X_{\alpha,\gamma}(t)$ and $Y_{\alpha,\gamma}(t)$. In the following,
we return to the case where $\langle \eta(t)\eta(t') \rangle
=\delta(t-t')$.

\section{Casimir Energy associated with $X_{\alpha,\gamma}(t)$ at Finite Temperature}

The ordinary oscillator process can be regarded as one-dimensional
Euclidean scalar massive field as its spectral density
$(\omega^2+\lambda^2)^{-1}$  is just the Euclidean propagator for a
scalar field with mass $\lambda$. By analogy, one can consider
$X_{\alpha,\gamma}(t)$ as a fractional Euclidean scalar massive
field in one-dimension with propagator
$(|\omega|^{2\alpha}+\lambda^2)^{-\gamma}$. From this viewpoint, it
will be interesting to find a convenient way to quantize the
corresponding fractional quantum field $\phi_{\alpha,\gamma}(t)$.
This can be achieved by using the stochastic quantization of Parisi
and Wu \cite{nr14}. According to this quantization scheme, an
additional auxiliary time $\tau$ is introduced and the Euclidean
quantum field $\phi_{\alpha,\gamma}(t;\tau)$ is assumed to evolve in
this auxiliary time according to a stochastic differential equation
of Langevin type with external white noise. The large equal-$\tau$
equilibrium limit of the
 covariance function of the solution to this
Langevin equation  gives the one-dimensional two-point Schwinger
function of the Euclidean field $\phi_{\alpha,\gamma}(t)$.

The nonlocal Euclidean action of the massive scalar field which
satisfies the (one-dimensional) fractional Klein-Gordon equation $$
\left[(-\Delta)^{\alpha}+\lambda^2\right]^{\gamma}\phi_{\alpha,\gamma}(t)=0$$
is given by \begin{align}\label{eq3_24_1}
S[\phi_{\alpha,\gamma}]=\frac{1}{2}\int\limits_{\R}
\phi_{\alpha,\gamma}(t)\left[(-\Delta)^{\alpha}+\lambda^2\right]^{\gamma}\phi_{\alpha,\gamma}(t)dt.
\end{align}  Parisi-Wu quantization procedure
requires $\phi_{\alpha,\gamma}(t;\tau)$  satisfying the following
stochastic differential equation
\begin{align}\label{eq3_24_2}
\frac{\pa \phi_{\alpha,\gamma}(t;\tau)}{\pa \tau}
=-\left.\frac{\delta S[\phi_{\alpha,\gamma}]}{\delta
\phi_{\alpha,\gamma}}\right|_{\phi_{\alpha,\gamma}=\phi_{\alpha,\gamma}(t;\tau)}+\eta(t;\tau),
\end{align} where $\eta(t,\tau)$  is the external white noise defined by
\begin{align*}
\left\langle \eta(t;\tau)\right\rangle =0,\hspace{1cm}\left\langle
\eta(t;\tau)\eta(t';\tau')\right\rangle
=2\delta(t-t')\delta(\tau-\tau').
\end{align*}
Eqs. \eqref{eq3_24_1} and \eqref{eq3_24_2} give
\begin{align}\label{eq3_24_3} \frac{\pa
\phi_{\alpha,\gamma}(t;\tau)}{\pa \tau}
=-\left[(-\Delta)^{\alpha}+\lambda^2\right]^{\gamma}\phi_{\alpha,\gamma}(t)+\eta(t;\tau).
\end{align} The solution of \eqref{eq3_24_3} subjected to the initial condition $\phi_{\alpha,
\gamma}(t;0)=0$  is \begin{align}\label{eq3_24_4}
\phi_{\alpha,\gamma}(t;\tau)=\int\limits_{\R} \int\limits_{0}^{\tau}
G(t-t'; \tau-\tau')\eta(t'; \tau')dt'd\tau',
\end{align} where $G(t;\tau)$ is the retarded Green's function given by
\begin{align*}
G(t;\tau)=\frac{\theta(\tau)}{2\pi}\int\limits_{\R}
\exp\left[-\left(|\omega|^{2\alpha}+\lambda^2\right)^{\gamma}\tau\right]e^{i\omega
t}d\omega.                                  \end{align*}The large
equal--$\tau$ limit ($\tau=\tau'\rightarrow \infty$) of the
  covariance function  gives
\begin{align}\label{eq3_25_1}
&\lim_{\tau_1=\tau_2\rightarrow \infty} \left\langle
\phi_{\alpha,\gamma}(t_1,\tau_1)\phi_{\alpha,\gamma}(t_2,\tau_2)\right\rangle
\\\nonumber =&2\lim_{\tau\rightarrow \infty}\int\limits_{\R}\int_{0}^{\tau}
G(t_1-t'; \tau-\tau')G(t_2-t'; \tau-\tau')d\tau'dt'\\\nonumber
=&\frac{1}{\pi}\int\limits_{\R}\int\limits_{0}^{\infty}\exp\left[-2\left(|\omega|^{2\alpha}+\lambda^2\right)^{\gamma}\tau'\right]
e^{i\omega(t_1-t_2)}d\tau' d\omega\\
=&\frac{1}{2\pi}\int\limits_{\R}\frac{e^{i\omega(t_1-t_2)}}{\left(|\omega|^{2\alpha}+\lambda^2\right)^{\gamma}}d\omega,\nonumber
\end{align} which is just the Euclidean two--point function of the fractional field $\phi_{\alpha,\gamma}(t)$.

We next consider the Parisi-Wu quantization method at positive
temperature $T=\beta^{-1}$. We follow the Matsubara imaginary time
formalism of finite temperature field theory by requiring
$\phi_{\alpha,\gamma}^T(t;\tau)$ to be periodic in the Euclidean
time $t$ with period $\beta$. That is, \begin{align}\label{eq3_25_2}
\phi_{\alpha,\gamma}^T(t+\beta;\tau)=\phi_{\alpha,\gamma}^T(t;\tau).
\end{align} In addition, the white noise $\eta^T(t;\tau)$ is assumed to satisfy the periodic
condition in $t$, i.e. $\eta^T(t+\beta; \tau)=\eta^T(t;\tau)$, such
that
\begin{align}\label{eq3_25_3}
\left\langle \eta^T(t;\tau)\right\rangle =0,\hspace{1cm}\left\langle
\eta^T(t;\tau)\eta^T(t';\tau')\right\rangle
=\frac{2}{\beta}\sum_{n=-\infty}^{\infty}\exp\left[i\omega_n(t-t')\right]\delta(\tau-\tau'),
\end{align}
 with $\omega_n =2\pi n/\beta$. The fractional operator
 $\left[(-\Delta)^{2\alpha}+\lambda^2\right]^{\gamma}$ acting on
 $\phi_{\alpha,\gamma}^T(t;\tau)$ is defined via Fourier series
 expansion (with respect to $t$)
of $\phi_{\alpha,\gamma}^T(t;\tau)$, i.e.,
\begin{align*}
\left[(-\Delta)^{2\alpha}+\lambda^2\right]^{\gamma}\phi_{\alpha,\gamma}^T(t;\tau)=
\frac{1}{\beta}\sum_{n=-\infty}^{n=\infty} \left[\int_0^{\beta}
\phi_{\alpha,\gamma}(t';\tau)e^{-i\omega_n
t'}dt'\right]\left(|\omega_n|^{2\alpha}+\lambda^2\right)^{\gamma}
e^{i\omega_n t}.
\end{align*}
 The retarded Green's function for the Langevin equation
\eqref{eq3_24_2}  satisfying the periodic conditions is given by
\begin{align}\label{eq3_25_4}
G^T(t;\tau)=\frac{\theta(\tau)}{\beta}\sum_{n=-\infty}^{\infty}
\exp\left[-\left(|\omega_n|^{2\alpha}+\lambda^2\right)^{\gamma}\tau\right]e^{i\omega_n
t}.
\end{align}
The solution with initial condition $\phi_{\alpha;\gamma}^T(t;0)=0$
is
\begin{align}\label{eq3_25_5}
\phi_{\alpha,\gamma}^T(t;\tau)=\int_{0}^{\beta}\int_{0}^{\tau}G^T(t-t';\tau-\tau')\eta^T(t';\tau')d\tau'dt',
\end{align}with
covariance given by
\begin{align}\label{eq3_25_6}
&\left\langle
\phi_{\alpha,\gamma}^T(t_1,\tau_1)\phi_{\alpha,\gamma}^T(t_2,\tau_2)\right\rangle=\frac{2
}{\beta}\sum_{n=-\infty}^{\infty}e^{i\omega_n(t_1-t_2)}\times\\&\nonumber\int_{0}^{\tau_1\wedge\tau_2}
\exp\left[-\left(|\omega_n|^{2\alpha}+\lambda^2\right)^{\gamma}(\tau_1-\tau')\right]
\exp\left[-\left(|\omega_n|^{2\alpha}+\lambda^2\right)^{\gamma}(\tau_2-\tau')\right]d\tau'
\end{align}
At the large  equal-$\tau$ limit, the covariance becomes
\begin{align} \label{eq3_25_7} \lim_{\tau_1=\tau_2\rightarrow \infty} \left\langle
\phi_{\alpha,\gamma}^T(t_1,\tau_1)\phi_{\alpha,\gamma}^T(t_2,\tau_2)\right\rangle
=\frac{1}{\beta}\sum_{n=-\infty}^{\infty}\frac{e^{i\omega_n(t_1-t_2)}}{\left(|\omega_n|^{2\alpha}+
\lambda^2\right)^{\gamma}},
\end{align} which is the thermal two-point function for the Euclidean fractional
Klein-Gordon field. When $\alpha=\gamma=1$,  this reduces to the
ordinary two-point function of one-dimensional Euclidean scalar
field at finite temperature \cite{Z}. We would also like to mention
that when $\beta\rightarrow \infty$, the limit of the thermal
two-point function \eqref{eq3_25_7} is the two point function
\eqref{eq3_25_1}.

We remark that  we can also consider the solution
$\phi_{\alpha,\gamma}^{\infty}(t;\tau)$ to \eqref{eq3_24_2}
satisfying the initial condition
$\phi_{\alpha,\gamma}^{\infty}(t;-\infty)=0$, i.e., instead of
having the field evolves from $\tau=0$, we require it to evolve from
$\tau=-\infty$. The solution $\phi_{\alpha,\gamma}^{\infty}(t;\tau)$
is then given by\begin{align}\label{eq3_24_4}
\phi_{\alpha,\gamma}(t;\tau)=\int\limits_{\R}
\int\limits_{-\infty}^{\tau} G(t-t'; \tau-\tau')\eta(t';
\tau')dt'd\tau'.
\end{align}It can be shown that
$\phi_{\alpha,\gamma}^{\infty}(t;\tau)$ is a stationary field and in
the large $\tau$ limit, the field $\phi_{\alpha,\gamma}(t;\tau)$
approaches the field $\phi_{\alpha,\gamma}^{\infty}(t;\tau)$, i.e.,
$$\lim_{\tau\rightarrow
\infty}\phi_{\alpha,\gamma}(t;\tau)=\phi_{\alpha,\gamma}^{\infty}(t;\tau).$$The
equal-$\tau$ variance of $\phi_{\alpha,\gamma}(t;\tau)$ is
independent of $\tau$ and is precisely the propagator
\eqref{eq3_25_1}. The same statement applies to
$\phi_{\alpha,\gamma}^{T,\infty}(t;\tau)$ which is the solution of
the periodic version of \eqref{eq3_24_2} with boundary condition
$\phi_{\alpha,\gamma}^{T,\infty}(t;-\infty)=0$.

We would also like to remark that just as the stochastic process
$X_{\alpha,\beta}(t)$ is related to the field
$\phi_{\alpha,\beta}(t)$ in the sense that the covariance function
of $X_{\alpha,\beta}(t)$ coincides with the propagator of
$\phi_{\alpha,\beta}(t)$, we can define a periodic stochastic
process $X_{\alpha,\beta}^T(t)$ whose covariance function is the
propagator of the periodic field $\phi_{\alpha,\gamma}^T(t)$
\eqref{eq3_25_7}. In fact, consider the solution of the fractional
stochastic differential equation
\begin{align*}
\left[(-\Delta)^{\alpha}+\lambda^2\right]^{\frac{\gamma}{2}}X_{\alpha,\gamma}^T(t)=\eta^T(t),
\end{align*}where $\eta^T(t)$ is the periodic white noise with
period $\beta$ and
\begin{align*}
\langle \eta^T(t)\rangle =0, \hspace{1cm}\left\langle
\eta^T(t)\eta^T(t')\right\rangle=\frac{1}{\beta}\sum_{n=-\infty}^{\infty}
e^{i\omega_n(t-t')}.
\end{align*}Using Fourier series, it is easy to check that the
solution is given by
\begin{align*}
X_{\alpha,\beta}^T(t)=\frac{1}{\beta}\sum_{n=-\infty}^{\infty}\int_0^{\beta}
\frac{e^{i\omega_n(t-t')}}{\left(|\omega_n|^{2\alpha}+\lambda^2\right)^{\frac{\gamma}{2}}}\eta(t')dt';
\end{align*}and the covariance function is
\begin{align*}
\left\langle X_{\alpha,\gamma}^T(t)
X_{\alpha,\gamma}^T(t')\right\rangle
=\frac{1}{\beta}\sum_{n=-\infty}^{\infty}\frac{e^{i\omega_n(t_1-t_2)}}{\left(|\omega_n|^{2\alpha}+
\lambda^2\right)^{\gamma}},
\end{align*}which coincides with the thermal two point function
\eqref{eq3_25_7}.

 Now we proceed to calculate the
partition function for the fractional oscillator process or
fractional Euclidean field in one-dimension at finite temperature,
hence its associated Casimir free energy. For this purpose, we
employ the technique of zeta function regularization \cite{a1, E1,
ET, K}. Due to the fractional character of the scalar field under
consideration, the derivation of Casimir free energy is more
complicated as compared with ordinary scalar field.

By definition, the Casimir free energy $\mathcal{F}$ of the
fractional Klein--Gordon field $\phi_{\alpha,\gamma}^T(t)$ which is
kept at thermal equilibrium with temperature $T=\beta^{-1}$ is given
by
$$\mathcal{F}=-\frac{1}{\beta}\log \mathcal{Z},$$where $\mathcal{Z}$
is
 the
partition function  defined by
\begin{align*}
\mathcal{Z}=\int
\mathcal{D}\phi^T\exp\left(-\frac{1}{2}\int_{0}^{\beta}\phi^T(t)
\left[\left(-\Delta\right)^{\alpha}+m^2\right]^{\gamma}\phi^T(t)dt\right).
\end{align*}
Using zeta regularization techniques, we find that
\begin{align}\label{eq3_25_10}
\mathcal{F}=-\frac{1}{2\beta}\left(\zeta'(0)-\zeta(0)\log
\mu^2\right),
\end{align}where $\mu$ is a normalization constant and $\zeta(s)$ is the zeta function
\begin{align}\label{eq3_11_2}
\zeta(s)=m^{-2\gamma
s}+2\sum_{n=1}^{\infty}\left\{[an]^{2\alpha}+m^2\right\}^{-\gamma
s},
\end{align}
with $a=\frac{2\pi}{\beta}$. The series  in \eqref{eq3_11_2} is
divergent when $s\leq 1/(2\alpha\gamma)$. Therefore we need to find
an analytic continuation of $\zeta(s)$ to a neighbourhood of $s=0$.
For this purpose, we use standard techniques and write
\begin{align}\label{eq3_25_8}
2\sum_{n=1}^{\infty}\left\{[an]^{2\alpha}+m^2\right\}^{-\gamma s}=&
\frac{2}{\Gamma(\gamma s)}\int_0^{\infty} t^{\gamma
s-1}\sum_{n=1}^{\infty} e^{-t\left([an]^{2\alpha}+m^2\right)}dt.
\end{align}Since the obstacle for this integral to define an analytic function in $s$ comes from
the singularity at $t=0$ of the integrand, the asymptotic behavior
of
\begin{align*}
2\sum_{n=1}^{\infty} e^{-t[an]^{2\alpha}}
\end{align*}as $t\rightarrow 0$ becomes crucial here. Using the representation
\begin{align}\label{eq3_11_7}
e^{-z}=\frac{1}{2\pi i}\int_{c-i\infty}^{c+i\infty} \Gamma(w)
z^{-w}dw,\hspace{1cm}c\in\R^+,
\end{align}we have
\begin{align*}
2\sum_{n=1}^{\infty} e^{-t[an]^{2\alpha}}= \frac{2}{2\pi
i}\int_{c-i\infty}^{c+i\infty} dw \Gamma(w) t^{-w} a^{-2\alpha w}
\zeta_R(2\alpha w),
\end{align*}when $c>1/(2\alpha)$. Here $\zeta_R(s)$ is the Riemann zeta function. This gives the asymptotic behavior
\begin{align}\label{eq3_11_3}
\sum_{n=1}^{\infty} e^{-t[an]^{2\alpha}}\sim
\frac{1}{\alpha}\Gamma\left(\frac{1}{2\alpha}\right)t^{-\frac{1}{2\alpha}}a^{-1}+2\sum_{k=0}^{\infty}
\frac{(-1)^k}{k!}t^k a^{2\alpha k}\zeta_R(-2\alpha k)
\end{align}as $t\rightarrow 0$ (also as $a\rightarrow 0$). Using the fact that $\zeta_R(0)=-1/2$,
then with
\begin{align*}
K(t)=2\sum_{n=1}^{\infty} e^{-t[an]^{2\alpha}}-\frac{1}{\alpha}
\Gamma\left(\frac{1}{2\alpha}\right)t^{-\frac{1}{2\alpha}}a^{-1}+1,
\end{align*}eq. \eqref{eq3_11_3} implies that $K(t)=O(t)$ as $t\rightarrow 0$. Now, we can
continue the evaluation of the integral in \eqref{eq3_25_8}:
\begin{align*}
&\frac{2}{\Gamma(\gamma s)}\int_0^{\infty} t^{\gamma
s-1}\sum_{n=1}^{\infty} e^{-t\left([an]^{2\alpha}+m^2\right)}dt\\
=&\frac{1}{\Gamma(\gamma s)}\int_0^{\infty} t^{\gamma
s-1}\left[\frac{1}{\alpha}\Gamma\left(\frac{1}{2\alpha}\right)t^{-\frac{1}{2\alpha}}a^{-1}-1\right]e^{-tm^2}dt+
\frac{1}{\Gamma(\gamma s)}\int_0^{\infty} t^{\gamma
s-1}K(t)e^{-tm^2}dt\\
=&\frac{1}{\alpha}\frac{\Gamma\left(\frac{1}{2\alpha}\right)\Gamma\left(\gamma
s-\frac{1}{2\alpha}\right)}{\Gamma(\gamma s)}a^{-1} m^{-2\gamma
s+\frac{1}{\alpha}} - m^{-2\gamma s}+\frac{1}{\Gamma(\gamma
s)}\int_0^{\infty} t^{\gamma s-1}K(t)e^{-tm^2}dt.
\end{align*}Since $K(t)=O(t)$ as $t\rightarrow 0$, the second integral in the last line
of the above equation defines an analytic function for
$s>-1/\gamma$. Combining with the first term in \eqref{eq3_11_2}, we
find that an analytic continuation of $\zeta(s)$ to
$\text{Re}\;s>-1/\gamma$ is given by
\begin{align}\label{eq3_25_9}
\zeta(s)=\frac{\Gamma\left(\frac{1}{2\alpha}\right)\Gamma\left(\gamma
s-\frac{1}{2\alpha}\right)}{\alpha\Gamma(\gamma s)}a^{-1}
m^{-2\gamma s+\frac{1}{\alpha}} +\frac{1}{\Gamma(\gamma
s)}\int_0^{\infty} t^{\gamma s-1}K(t)e^{-tm^2}dt.
\end{align}To evaluate $\zeta(0)$ and $\zeta'(0)$, we observe that$$\int_0^{\infty} t^{\gamma s-1}K(t)e^{-tm^2}dt$$
is analytic for $s>-1/\gamma$. Therefore the only possible
contribution to $\zeta(0)$ comes from the first term in
\eqref{eq3_25_9} when $1/(2\alpha)\in \mathbb{N}$. Denote by
$\Lambda$ the set
$$\Lambda=\left\{\frac{1}{2u}\,:\,u\in \mathbb{N}\right\},$$and let
$$\omega_{\alpha,\Lambda}=\begin{cases}
1, \hspace{0.5cm}\text{if}\;\; \alpha\in\Lambda\\
0,\hspace{0.5cm}\text{if}\;\;\alpha\notin\Lambda\end{cases}.$$ Then
we find that
\begin{align*}
\zeta(0)=2\omega_{\alpha,\Lambda}
(-1)^{\frac{1}{2\alpha}}a^{-1}m^{\frac{1}{\alpha}},
\end{align*}and
\begin{align*}
\zeta'(0)=&2\omega_{\alpha,\Lambda}
(-1)^{\frac{1}{2\alpha}}a^{-1}m^{\frac{1}{\alpha}}\gamma\left\{\psi\left(\frac{1}{2\alpha}+1\right)-\psi(1)
-\log m^2\right\}\\
&-
\gamma(1-\omega_{\alpha,\Lambda})\frac{2\pi}{\sin\frac{\pi}{2\alpha}}a^{-1}m^{\frac{1}{\alpha}}
+\gamma\int_0^{\infty}t^{-1}K(t)e^{-tm^2}dt.
\end{align*}Here $\psi(z)=\Gamma'(z)/\Gamma(z)$ is the logarithmic derivative of the gamma
function. Substituting the above into \eqref{eq3_25_10} gives the
Casimir free energy
\begin{align*}
\mathcal{F}=&\frac{1}{2\beta}\Biggl\{\omega_{\alpha,\Lambda}
(-1)^{\frac{1}{2\alpha}}\frac{\beta}{\pi}m^{\frac{1}{\alpha}}\left[\log\mu^2-\gamma\left(\psi\left(\frac{1}{2\alpha}
+1\right)-\psi(1)
-\log m^2 \right)\right]\\
&+
\gamma(1-\omega_{\alpha,\Lambda})\frac{\beta}{\sin\frac{\pi}{2\alpha}}m^{\frac{1}{\alpha}}
-\gamma\int_0^{\infty}t^{-1}K(t)e^{-tm^2}dt\Biggr\}.
\end{align*}To study the asymptotic behavior of $\mathcal{F}$,
we first use \eqref{eq3_11_3} to obtain an asymptotic behavior of
$\zeta(s)$ when $a\rightarrow 0$:
\begin{align*}
\zeta(s)\sim\frac{\Gamma\left(\frac{1}{2\alpha}\right)\Gamma\left(\gamma
s-\frac{1}{2\alpha}\right)}{\alpha\Gamma(\gamma s)}a^{-1}
m^{-2\gamma s+\frac{1}{\alpha}}+2\sum_{k=1}^{\infty}
\frac{(-1)^k}{k!}\frac{\Gamma(\gamma s+k)}{\Gamma(\gamma
s)}m^{-2\gamma s-2k} a^{2\alpha k}\zeta_R(-2\alpha k).
\end{align*}From this we can find the asymptotic behavior of $\zeta(0)$ and
$\zeta'(0)$, which, when substituted into \eqref{eq3_25_10} gives
\begin{align}\label{eq3_14_3}
\mathcal{F}\sim &\omega_{\alpha,\Lambda}\frac{
(-1)^{\frac{1}{2\alpha}}}{2\pi}m^{\frac{1}{\alpha}}\left[\log\mu^2-\gamma\left(\psi\left(\frac{1}{2\alpha}
+1\right)-\psi(1)
-\log m^2\right)\right]\\
&+\gamma(1-\omega_{\alpha,\Lambda})\frac{m^{\frac{1}{\alpha}}}{2\sin\frac{\pi}{2\alpha}}
-\gamma \sum_{k=1}^{\infty} \frac{(-1)^k}{k}m^{-2k} (2\pi)^{2\alpha
k}\zeta_R(-2\alpha k)\beta^{-2\alpha k-1}.\nonumber
\end{align}when $\beta\rightarrow \infty$. In the special case $\alpha=1$, since $\zeta(-2k)=0$ for all
$k\in\mathbb{N}$, \eqref{eq3_14_3} gives us
\begin{align}\label{eq3_11_4}
\mathcal{F}\sim\frac{\gamma m}{2}.
\end{align}In fact, by applying the Jacobi inversion formula
\begin{align*}
1+2\sum_{n=1}^{\infty}e^{-t[an]^2}=\frac{1}{a}\sqrt{\frac{\pi}{t}}\sum_{n=-\infty}^{\infty}
e^{-\frac{\pi^2}{ta^2}n^2},
\end{align*}we find that when $\alpha=1$,
\begin{align*}
\zeta(s)=&\frac{\sqrt{\pi}}{a\Gamma(\gamma s)}\int_0^{\infty}
t^{\gamma
s-\frac{1}{2}-1}\sum_{n=-\infty}^{\infty}e^{-\frac{\pi^2}{ta^2}n^2
-tm^2}dt\\
=&\frac{\sqrt{\pi}\Gamma\left(\gamma
s-\frac{1}{2}\right)}{a\Gamma(\gamma s)}m^{1-2\gamma
s}+\frac{4\sqrt{\pi}}{a\Gamma(\gamma
s)}\sum_{n=1}^{\infty}\left(\frac{\pi n}{am}\right)^{\gamma
s-\frac{1}{2}}K_{\gamma s-\frac{1}{2}}\left(\frac{2\pi n
m}{a}\right).
\end{align*}Here $K_{\nu}(z)$ is the modified Bessel function of second kind. Together with $K_{1/2}(z)=\sqrt{
\pi/(2z)}e^{-z}$, one obtains
\begin{align}\label{eq3_11_5}
\mathcal{F} = \frac{\gamma
m}{2}-\frac{\gamma}{\beta}\sum_{n=1}^{\infty}\frac{1}{n}e^{-\beta
nm}=\frac{\gamma m}{2}+\frac{\gamma}{\beta}\log\left(1-e^{-\beta
m}\right)=\frac{\gamma}{\beta}\log \left[2\sinh\frac{\beta
m}{2}\right].
\end{align}
It is easy to see that the leading term when $\beta\gg 1$ agrees
with \eqref{eq3_11_4} and the remainder terms decay exponentially.

From \eqref{eq3_14_3}, we note that at low temperature $T\ll 1$
($\beta\gg 1$), the leading order term of the free energy
$\mathcal{F}$ is of order $T^0$. When $\alpha\notin \Lambda$, i.e.
$\alpha$ is not the reciprocal of an even number, then the leading
order is
\begin{align}\label{eq4_17_1}
\mathcal{F}\sim \frac{\gamma
m^{\frac{1}{\alpha}}}{2\sin\frac{\pi}{2\alpha}}+O(T^{1+2\alpha}),
\end{align}whose sign depends on $\alpha$.
There is a dependence of $\mathcal{F}$ on the normalization constant
$\mu$ when $\alpha\in \Lambda$. We shall renormalize the free energy
$\mathcal{F}$ to get rid of this dependence later.

To study the high temperature  behavior of $\mathcal{F}$, we can use
the expansion
\begin{align}\label{eq3_10_1}\exp(-tm^2)=\sum_{l=0}^{\infty}
\frac{(-1)^l}{l!}t^lm^{2l}\end{align} and find that
\begin{align}\label{eq3_11_1}
2\sum_{n=1}^{\infty}\left\{[an]^{2\alpha}+m^2\right\}^{-\gamma s}=&
\frac{2}{\Gamma(\gamma s)}\sum_{l=0}^{\infty}\frac{(-1)^l}{l!}
m^{2l}\int_0^{\infty} t^{\gamma s+l-1}\sum_{n=1}^{\infty}
e^{-t[an]^{2\alpha}}dt\\
=&\frac{2}{\Gamma(\gamma s)}\sum_{l=0}^{\infty}\frac{(-1)^l}{l!}
m^{2l}a^{-2\alpha(\gamma s+l)}\Gamma( \gamma
s+l)\zeta_R(2\alpha(\gamma s+l)).\nonumber
\end{align}The $l=0$ term in \eqref{eq3_11_1} will contribute $-1 $ to
$\zeta(0)$, which is canceled by the contribution from the term
$m^{-2\gamma s}$ \eqref{eq3_11_2}. Moreover, since $ \zeta_R(s)$ is
meromorphic on $\C$ with a simple pole at   $s=1$ with
\begin{align*} \zeta_R(s) =&
 \frac{1}{s-1}-\psi(1) +O\left(s-1\right),\hspace{1cm}\text{as}\;\;
s\rightarrow 1,\nonumber
\end{align*} we see
that  if $\alpha=1/(2j)$ for some $j\in\mathbb{N}$, then there is
another nonzero contribution to $\zeta(0)$ arising from the $l=j$
term in \eqref{eq3_11_1}. Therefore,
\begin{align*}
\zeta(0)=2\omega_{\alpha,\Lambda}
(-1)^{\frac{1}{2\alpha}}a^{-1}m^{\frac{1}{\alpha}},
\end{align*}and
\begin{align*}
\zeta'(0)=&-2\gamma\log
m-2\alpha\gamma\log\frac{2\pi}{a}+2\gamma\sum_{\substack{l\in\mathbb{N}\\l\neq
\frac{1}{2\alpha}}}\frac{(-1)^l}{l} m^{2l}a^{-2\alpha l}
\zeta_R(2\alpha l)\\&-2\omega_{\alpha,\Lambda}
(-1)^{\frac{1}{2\alpha}}a^{-1}m^{\frac{1}{\alpha}}\gamma\left(\alpha
\Bigl[\log a^2+2\psi(1)\Bigr]-  \psi\left(\frac{1}{2\alpha}
\right)+\psi(1) \right).
\end{align*}This gives us
\begin{align}\label{eq3_14_4}
\mathcal{F}=&\frac{1}{2\beta}\Biggl\{\gamma\log
m^2+2\alpha\gamma\log\beta-2\gamma\sum_{\substack{l\in\mathbb{N}\\l\neq
\frac{1}{2\alpha}}}\frac{(-1)^l}{l}
m^{2l}\left(\frac{\beta}{2\pi}\right)^{2\alpha l}\zeta_R(2\alpha
l)\\&+ \omega_{\alpha,\Lambda}
(-1)^{\frac{1}{2\alpha}}\frac{\beta}{\pi}m^{\frac{1}{\alpha}}
\left[\log\mu^2+\alpha\gamma\left( \log
\left(\frac{2\pi}{\beta}\right)^2+2\psi(1)\right)-\gamma\left[\psi\left(\frac{1}{2\alpha}
\right)-\psi(1)\right]\right]\Biggr\}.\nonumber
\end{align}In particular, in the high temperature limit $T\gg 1$ ($\beta\ll 1$),
\begin{align}\label{eq4_17_3}
\mathcal{F}\sim -\alpha\gamma T\log T+\frac{\gamma T}{2}\log
m^2+O(1),
\end{align}the Casimir energy is negative and the leading term
$-\alpha\gamma T\log T$ depends linearly on $\alpha$ and $\gamma$.
When $\alpha=1$, using
$$\zeta_R(2l)=(-1)^{l+1}\frac{(2\pi)^{2l}}{2(2l)!}B_{2l},$$ where
$B_{2l}$ is the Bernoulli numbers defined by
\begin{align*}
\frac{1}{e^{x}-1}=\frac{1}{x}
-\frac{1}{2}+\sum_{n=1}^{\infty}\frac{B_{2n}}{(2n)!}x^{2n-1};
\end{align*}and
\begin{align*}
\log
\frac{1-e^{-x}}{x}=\int_0^x\left\{\frac{1}{e^u-1}-\frac{1}{u}\right\}du=
-\frac{x}{2}+\frac{1}{2}\sum_{n=1}^{\infty}\frac{B_{2n}}{(2n)!}\frac{x^{2n}}{n},
\end{align*}we find that \eqref{eq3_14_4} gives
\begin{align*}
\mathcal{F}=&\frac{1}{2\beta}\left\{\gamma\log m^2+2\gamma\log
\beta+\gamma\sum_{l=1}^{\infty}\frac{B_{2l}}{(2l)!}\frac{(\beta
m)^{2l}}{l}\right\}\\
=&\frac{\gamma}{2\beta}\left\{2\log [\beta
m]+2\log\left[\frac{1-e^{-\beta m}}{\beta m}\right]+\beta
m\right\}=\frac{\gamma}{\beta}\log \left[2\sinh\frac{\beta
m}{2}\right],
\end{align*}agreeing with \eqref{eq3_11_5}.
\begin{figure}\centering \epsfxsize=.6\linewidth
\epsffile{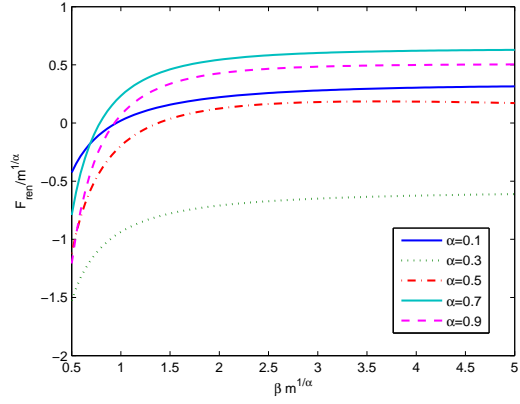}\caption{The renormalized free energy
$\mathcal{F}_{\text{ren}}/m^{\frac{1}{\alpha}}$ as a function of
$\beta m^{1/\alpha}$ when $\alpha=0.1, 0.3, 0.5, 0.7, 0.9$ and
$\gamma=1$}\end{figure}
\begin{figure}\centering \epsfxsize=.6\linewidth
\epsffile{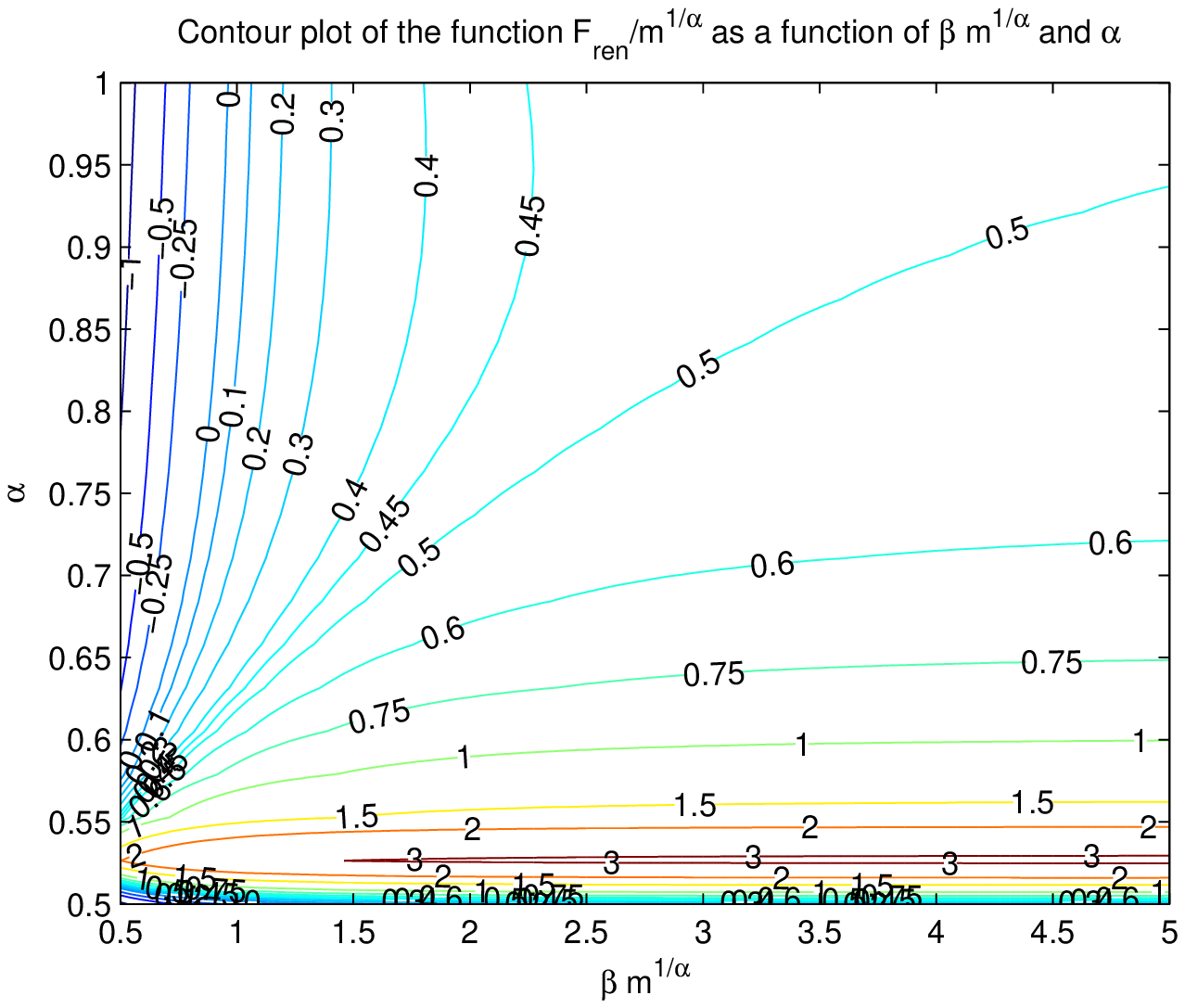}\caption{Contour plot of the renormalized free
energy $\mathcal{F}_{\text{ren}}/m^{\frac{1}{\alpha}}$ as a function
of $\beta m^{1/\alpha}$ and $\alpha$. Here $\gamma=1$.}\end{figure}

\begin{figure}\centering \epsfxsize=.8\linewidth
\epsffile{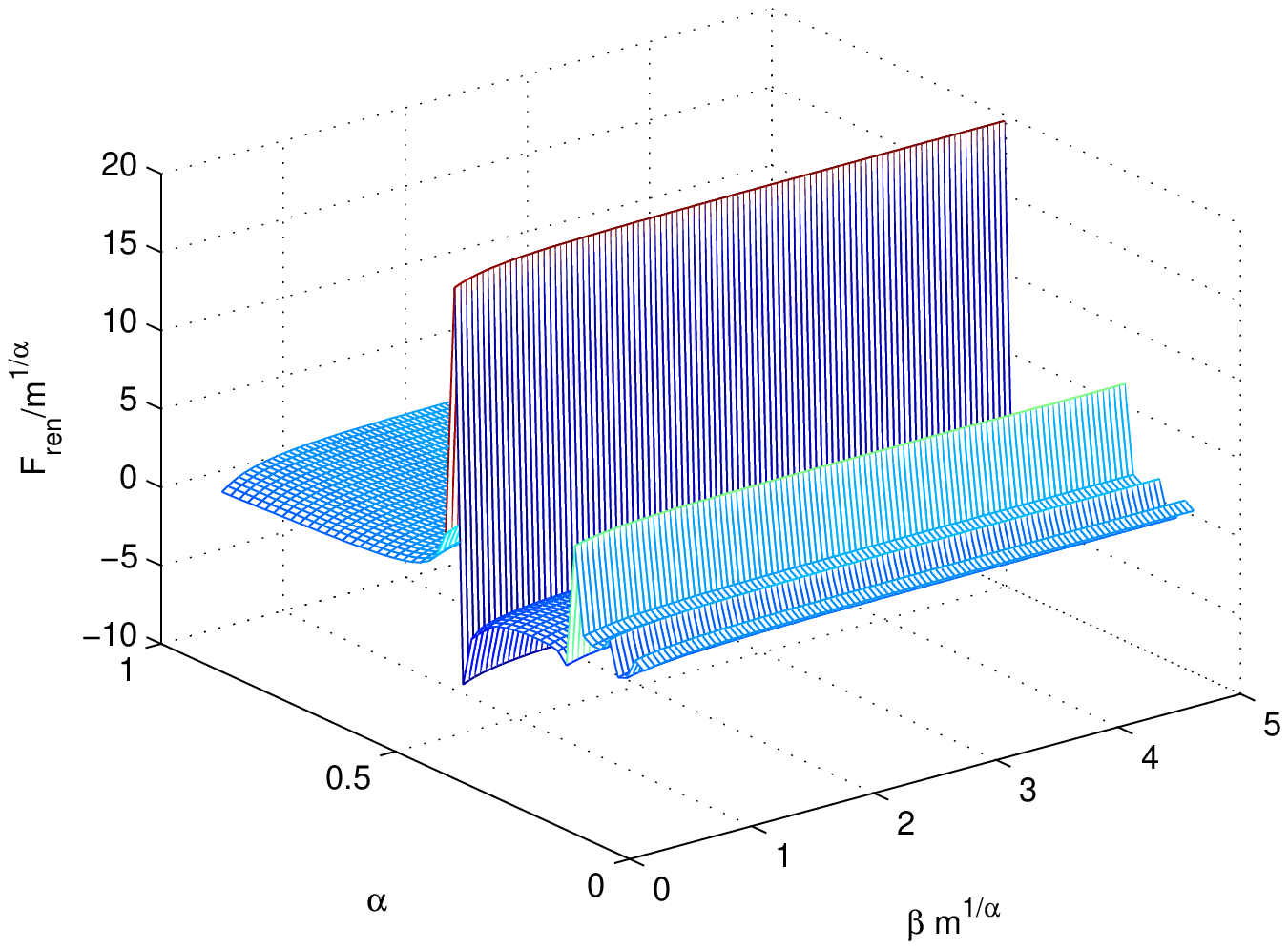}\caption{The renormalized free energy
$\mathcal{F}_{\text{ren}}/m^{\frac{1}{\alpha}}$ as a function of
$\beta m^{1/\alpha}$ and $\alpha$. Here $\gamma=1$.}\end{figure}

\begin{figure}\centering \epsfxsize=.52\linewidth
\epsffile{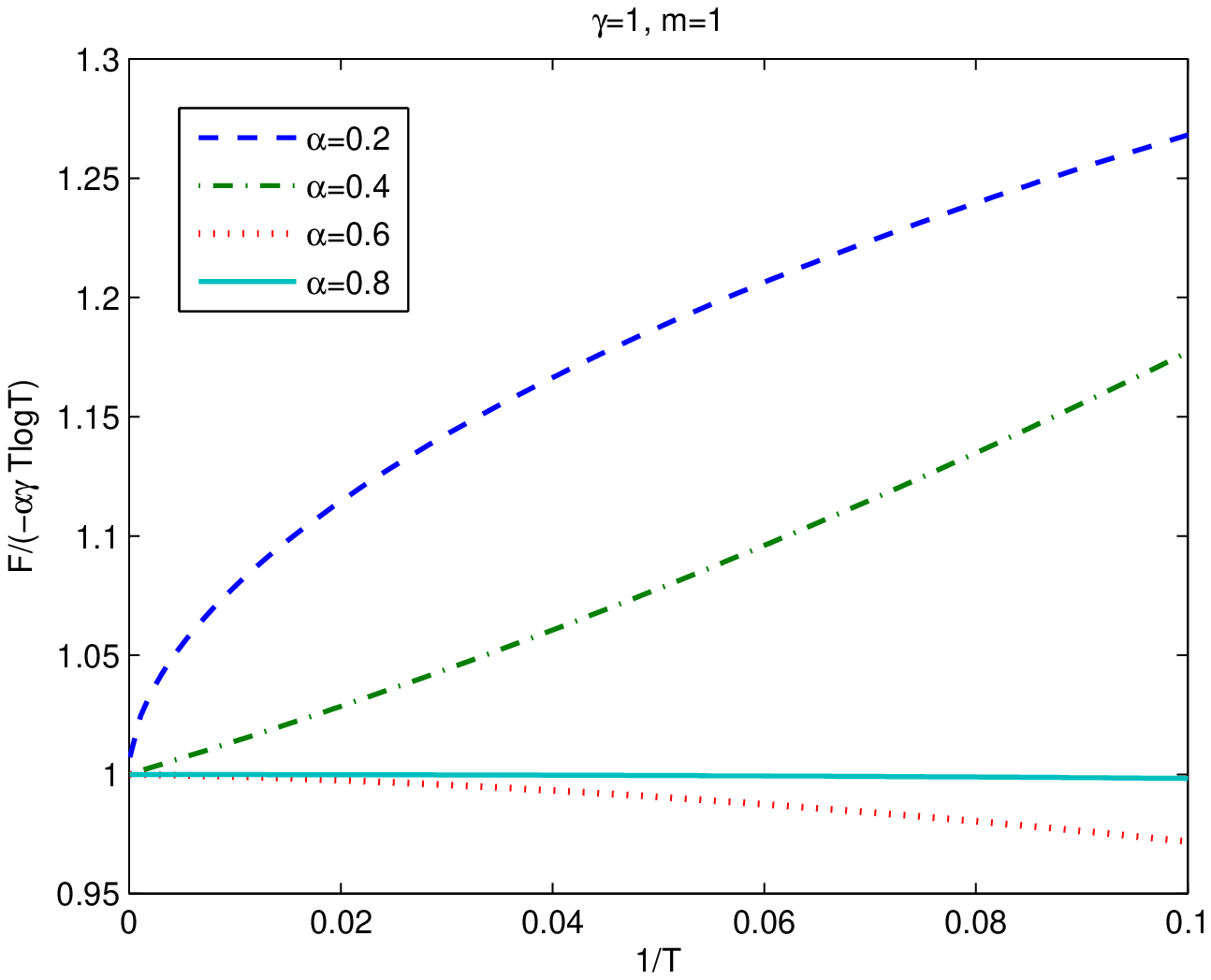}\centering \epsfxsize=.52\linewidth
\epsffile{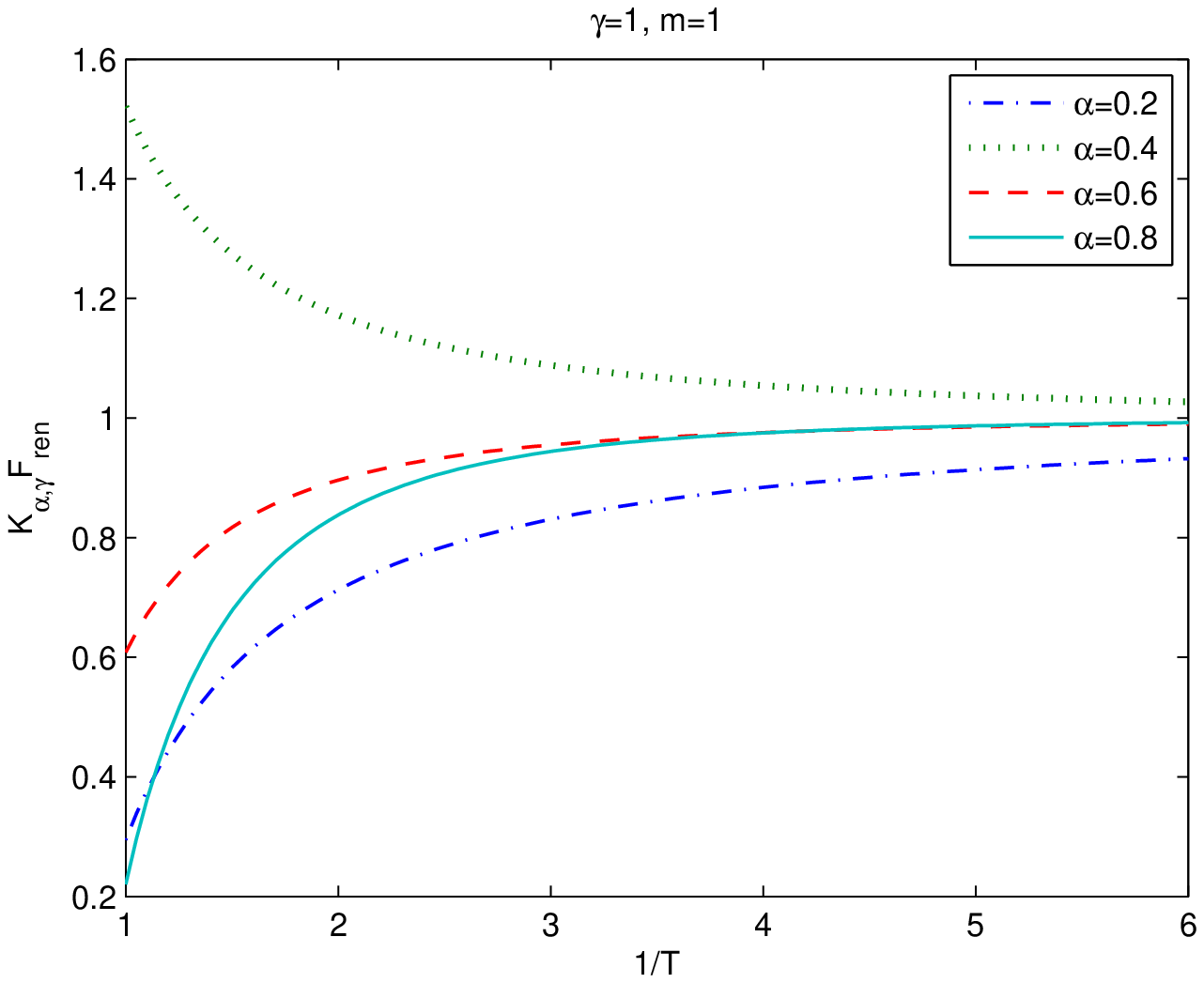}\caption{Left: This figure shows the ratio of
the renormalized free energy $\mathcal{F}_{\text{ren}}$  to
$-\alpha\gamma T\log T$ when $\beta=1/T$ is small. Here $\gamma=1$
and $m=1$. Right: This figure shows the small $T$ (large $\beta$)
behavior of the renormalized free energy $\mathcal{F}_{\text{ren}}$.
Here $K_{\alpha,\gamma}=2\sin\frac{\pi}{2\alpha}/(\gamma
m^{1/\alpha})$ and $\gamma=m=1$.}\end{figure}

One notices that when $\alpha\in \Lambda$, the free energy depends
on the normalization constant $\mu$. In order to remove this
dependence, we need to renormalize the free energy by adding a
counterterm $\mathcal{F}_c$ to the free energy so that the
renormalized free energy $\mathcal{F}_{\text{ren}}$ is
$$\mathcal{F}_{\text{ren}}=\mathcal{F}+\mathcal{F}_c.$$A reasonable
way to determine the counterterm $\mathcal{F}_c$ is to require that
in the limit $\beta\rightarrow \infty$ and $m\rightarrow 0$,
$\mathcal{F}_{\text{ren}}\rightarrow 0$. Eq. \eqref{eq3_14_3} gives
us immediately
$$\mathcal{F}_c=-\omega_{\alpha,\Lambda}\frac{(-1)^{\frac{1}{2\alpha}}}{2\pi}m^{\frac{1}{\alpha}}
\left(\log\mu^2+\gamma \log m^2\right).$$Note that adding the
counterterm $\mathcal{F}_c$ to the free energy is equivalent to
setting $\mu=m^{-\gamma}$ when $\alpha\in \Lambda$. Therefore, from
\eqref{eq3_14_3} and \eqref{eq3_14_4}, we obtain immediately that in
the low temperature $T=1/\beta\ll 1$ limit,
\begin{align}\label{eq3_17_1}
\mathcal{F}_{\text{ren}}\sim &-\gamma \omega_{\alpha,\Lambda}\frac{
(-1)^{\frac{1}{2\alpha}}}{2\pi}m^{\frac{1}{\alpha}}\left(\psi\left(\frac{1}{2\alpha}
+1\right)-\psi(1)
\right)\\
&+\gamma(1-\omega_{\alpha,\Lambda})\frac{m^{\frac{1}{\alpha}}}{2\sin\frac{\pi}{2\alpha}}
-\gamma \sum_{k=1}^{\infty} \frac{(-1)^k}{k}m^{-2k} (2\pi)^{2\alpha
k}\zeta_R(-2\alpha k)\beta^{-2\alpha k-1};\nonumber
\end{align}whereas in the high temperature  $T=1/\beta\gg 1$ limit,
\begin{align}\label{eq3_17_2}
\mathcal{F}_{\text{ren}}=&\frac{1}{2\beta}\Biggl\{\gamma\log
m^2+2\alpha\gamma\log\beta-2\gamma\sum_{\substack{l\in\mathbb{N}\\l\neq
\frac{1}{2\alpha}}}\frac{(-1)^l}{l}
m^{2l}\left(\frac{\beta}{2\pi}\right)^{2\alpha l}\zeta_R(2\alpha
l)\\&+\gamma \omega_{\alpha,\Lambda}
(-1)^{\frac{1}{2\alpha}}\frac{\beta}{\pi}m^{\frac{1}{\alpha}}
\left[\alpha\left( \log \left(\frac{2\pi}{\beta
m^{\frac{1}{\alpha}}}\right)^2+2\psi(1)\right)-\psi\left(\frac{1}{2\alpha}
\right)+\psi(1)\right]\Biggr\}.\nonumber
\end{align}
Notice that \eqref{eq3_17_1} implies that if $\alpha\in \Lambda$,
then when $T\rightarrow 0$, the leading order term of
$\mathcal{F}_{\text{ren}}$ is
\begin{align}\label{eq4_17_2}
\mathcal{F}_{\text{ren}}\sim -\gamma\frac{
(-1)^{\frac{1}{2\alpha}}}{2\pi}m^{\frac{1}{\alpha}}\left(\psi\left(\frac{1}{2\alpha}
+1\right)-\psi(1) \right)+O(T^{2\alpha+1}),
\end{align}whose sign depends on $\alpha$.

From \eqref{eq4_17_1} and \eqref{eq4_17_2}, we see that in the low
temperature limit, the leading order term of the renormalized
Casimir free energy still depends  linearly in $\gamma$, but its
dependence on $\alpha$ is highly nontrivial.

Finally, we observe from \eqref{eq3_17_1} and \eqref{eq3_17_2} that
the combination $\mathcal{F}_{\text{ren}}/m^{\frac{1}{\alpha}}$
depends on $\beta$ and $m$ in the combination $\beta
m^{\frac{1}{\alpha}}$. The graphs and contour plots of
$\mathcal{F}_{\text{ren}}/m^{\frac{1}{\alpha}}$ as a function of
$\alpha$ and $\beta m^{\frac{1}{\alpha}}$ are shown in Figures 7, 8
and 9. The high temperature and low temperature behaviors of
$\mathcal{F}_{\text{ren}}$ are shown in Figure 10.

\section{Concluding Remarks}
We have introduced a new Gaussian process called the fractional
oscillator process with two indices, which is obtained as the
solution to a stochastic differential equation with two fractional
orders. Some basic properties of this process can be obtained based
on the asymptotic properties of its covariance despite its complex
nature. The possible generalizations of fluctuation--dissipation
theorem and Einstein relation to this fractional process are
considered. The main advantage of the oscillator process
parametrized by two indices over the process with single index is
that the former has its fractal (or local self-similarity) and the
short range dependence separately characterized by the two indices,
while the later has both these properties determined by a single
index. Such a process may provide a more flexible model for
applications in phenomena with short memory.

    By analogy regarding the fractional oscillator process as the Euclidean
    fractional scalar Klein-Gordon field in one dimension, we carry out
    the stochastic quantization of such a field with a nonlocal action. The Casimir energy
    associated with the fractional Klein-Gordon field at finite positive
    temperature was calculated using thermal zeta function regularization
    technique. The expression for the free energy has a rather complicated form.
     We thus consider the low and high temperature limits for the free energy.
     Graphical representations of these asymptotic behaviors of the free energy are given.

Extension of our results  can be carried out to give
    the $n$-dimensional Euclidean fractional Klein-Gordon field. However, the derivation of
    the asymptotic  properties for the covariance function and the
Casimir energy will be more complicated. The sign dependency of the
free energy on $\alpha$  has important physical implication when the
fractional quantum field under consideration is confined between
parallal plates or cavities as the sign of Casimir energy will
determine whether the associated Casimir force is attractive or
repulsive.

    Another direction of
generalization is that of fractional oscillator process of variable
order, with $\alpha$ and $\gamma$ being extended to time-dependent
$\alpha(t)$ and $\gamma(t)$. We expect the variable short range
dependence property remains valid, and the result for fractal
dimension holds only locally. However, the Casimir energy
calculation will require new mathematical techniques and
approximations. Such a generalization may find applications for
complex systems where the physical phenomena can have variable short
memory and the fractal dimension varies with time or position.

\vspace{1cm} \noindent \textbf{Acknowledgement}\; The authors would
like to thank Malaysian Academy of Sciences, Ministry of Science,
Technology  and Innovation for funding this project under the
Scientific Advancement Fund Allocation (SAGA) Ref. No P96c.

\appendix \section{Mathematical details}

\noindent 1. We show that if the position process
$Y_{\alpha,\gamma}(t)$ is related to the velocity process
$X_{\alpha,\gamma}(t)$ by
$$X_{\alpha,\gamma}(t)=\frac{dY_{\alpha,\gamma}(t)}{dt},$$then the long time behavior
of the covariance function $C_{\alpha,\gamma}(\tau)$ of
$X_{\alpha,\gamma}(t)$ does not show up in the leading term in the
long time asymptotic expression of the variance $\left\langle
\left[Y_{\alpha,\gamma}(t)\right]^2\right\rangle$. Its effect only
appears in the second leading term. In fact, from \eqref{eq4_17_8},
we have
\begin{align*}
\left\langle \left[Y_{\alpha,\gamma}(t)\right]^2\right\rangle
=2\int_0^t (t-\tau)C_{\alpha,\gamma}(\tau)d\tau=2\left[\int_0^{t}
C_{\alpha,\gamma}(\tau) d\tau\right] t  -2\int_0^{t} \tau
C_{\alpha,\gamma}(\tau) d\tau.
\end{align*} Eq. \eqref{eq12_20_3} gives us the long-time behavior of
$C_{\alpha,\gamma}(\tau)$ as $C_{\alpha,\gamma}(\tau)\sim
A_1\tau^{-2\alpha-1}+O(\tau^{-4\alpha-1})$ for some constant $A_1$,
which implies that the integrals
$\int_0^{\infty}C_{\alpha,\gamma}(\tau)d\tau$ is convergent; and
\begin{align}\label{eq4_17_14}
\int_0^t C_{\alpha,\gamma}(\tau) d\tau
=\int_0^{\infty}C_{\alpha,\gamma}(\tau)d\tau-\int_t^{\infty}
C_{\alpha,\gamma}(\tau) d\tau & \sim A_2+
A_3t^{-2\alpha}+O(t^{-4\alpha})
\end{align}as $t\rightarrow \infty$. On the other hand, the
integral $\int_0^{\infty} \tau C_{\alpha,\gamma}(\tau) d\tau$ is
convergent if and only if $\alpha>1/2$. In this case
\begin{align}\label{eq4_17_11}\int_0^t \tau C_{\alpha,\gamma}(\tau)
d\tau =\int_0^{\infty} \tau C_{\alpha,\gamma}(\tau) d\tau
-\int_t^{\infty} \tau C_{\alpha,\gamma}(\tau) d\tau \sim A_4+
O(t^{-2\alpha+1})\end{align} as $t\rightarrow \infty$.
 If $\alpha<1/2$, then
\begin{align}\label{eq4_17_12} \int_0^{t} \tau C(\tau) d\tau \sim A_5 t^{1-2\alpha}
+O(t^{\max\{0, 1-4\alpha\}})
\end{align}as $t\rightarrow \infty$. In the borderline case
$\alpha=1/2$,
\begin{align}\label{eq4_17_13} \int_0^{t} \tau C(\tau) d\tau \sim A_6 \log t
+O(1).
\end{align}
Therefore, as $t\rightarrow \infty$, if $\alpha>1/2$, then
\begin{align*}
\left\langle \left[Y_{\alpha,\gamma}(t)\right]^2\right\rangle
 \sim & [2A_2] t  -[2A_4]+O( t^{1-2\alpha}).
\end{align*}
If $\alpha=1/2$, then
\begin{align*}
\left\langle \left[Y_{\alpha,\gamma}(t)\right]^2\right\rangle \sim
[2A_2] t -[2A_6] \log t+O(1).\end{align*}Finally, if $\alpha<1/2$,
then
\begin{align*}
\left\langle \left[Y_{\alpha,\gamma}(t)\right]^2\right\rangle \sim
[2A_2] t +[2A_3-2A_5] t^{1-2\alpha}+O(t^{\max\{0,
1-4\alpha\}}).\end{align*} These show that the leading term of
$\left\langle \left[Y_{\alpha,\gamma}(t)\right]^2\right\rangle$ is
of order $t$, independent of $\alpha$; and the second leading term
is of order $t^{\max\{0, 1-2\alpha\}}\log t$, which depends on
$\alpha$.

\vspace{0.5cm}\noindent 2. We show that if the position process
$Y_{\alpha,\gamma}(t)$ is related to the velocity process
$X_{\alpha,\gamma}(t)$ by \begin{align*}
X_{\alpha,\gamma}(t)=\;_0D_t^{\chi}Y_{\alpha,\gamma}(t),\hspace{1cm}
\frac{1}{2}<\chi <\frac{3}{2},\end{align*} and
$\left._0D_{t}^{\chi-j}Y_{\alpha,\gamma}(t)\right|_{t=0}=0$ for
$j=1$ if $\chi\leq 1$ and $j=1,2$ if $\chi>1$, then
\begin{align}\label{eq4_16_7_1}
\left\langle
\left[Y_{\alpha,\gamma}(t)\right]^2\right\rangle=2B\lambda^{-2\alpha\gamma}\left[\frac{t^{2\chi-1}}{
(2\chi-1) \Gamma(\chi)^2}\right]+O\left(t^{\max\{0, 2\chi-2,
2\chi-2\alpha-1\}}\log t\right).\end{align} From \eqref{eq4_17_9},
we have
\begin{align*}
\left\langle \left[Y_{\alpha,\gamma}(t)\right]^2\right\rangle =&
\frac{1}{\Gamma(\chi)^2}\int_0^t \int_0^t
(t-s_1)^{\chi-1}(t-s_2)^{\chi-1} \langle
X_{\alpha,\gamma}(s_1)X_{\alpha,\gamma}(s_2)\rangle ds_1 ds_2\\
=& \frac{1}{\Gamma(\chi)^2}\int_0^t \int_0^t
(t-s_1)^{\chi-1}(t-s_2)^{\chi-1} C_{\alpha,\gamma}(|s_1-s_2|)ds_1
ds_2.
\end{align*} Using some calculus, this gives
\begin{align*}
\left\langle \left[Y_{\alpha,\gamma}(t)\right]^2\right\rangle
=&\frac{2}{\Gamma(\chi)^2}\int_0^t \int_0^{s_2}
(t-s_1)^{\chi-1}(t-s_2)^{\chi-1} C_{\alpha,\gamma}(s_2-s_1) ds_1
ds_2\\
=&\frac{2}{\Gamma(\chi)^2}\int_0^t\int_0^s
(t+\tau-s)^{\chi-1} (t-s)^{\chi-1} C_{\alpha,\gamma}(\tau)d\tau ds\\
=&\frac{2}{\Gamma(\chi)^2}\int_0^t\left[
\int_{\tau}^t(t+\tau-s)^{\chi-1}
(t-s)^{\chi-1}ds\right] C_{\alpha,\gamma}(\tau)d\tau\\
=&\frac{2}{\Gamma(\chi)^2}\int_0^t\left[ \int_{\tau}^t u^{\chi-1}
(u-\tau)^{\chi-1} du\right] C_{\alpha,\gamma}(\tau)d\tau.
\end{align*}The case $\chi=1$ has been considered above. Now we consider  the cases $\chi\in (1/2, 1)$
and $\chi\in (1,3/2)$ separately. If $\chi\in (1/2, 1)$, using
\begin{align*}
(u-\tau)^{\chi-1}=u^{\chi-1}
-\left(u^{\chi-1}-(u-\tau)^{\chi-1}\right),
\end{align*}
we have
\begin{align*}
\int_\tau^t u^{\chi-1}(u-\tau)^{\chi-1}du=
\frac{1}{2\chi-1}\left(t^{2\chi-1}-\tau^{2\chi-1}\right)-\int_\tau^t
u^{\chi-1}\left(u^{\chi-1}-(u-\tau)^{\chi-1}\right)du.
\end{align*}
Therefore, if $\chi\in (1/2, 1)$,
\begin{align*}
\left\langle
[Y_{\alpha,\gamma}(t)]^2\right\rangle=&\frac{2}{\Gamma(\chi)^2}\Biggl(
\frac{t^{2\chi-1}}{2\chi-1}\int_0^t
C_{\alpha,\gamma}(\tau)d\tau-\frac{1}{2\chi-1}\int_0^t
\tau^{2\chi-1}C_{\alpha,\gamma}(\tau) d\tau \\&-\int_0^t
\left[\int_\tau^t
u^{\chi-1}\left(u^{\chi-1}-(u-\tau)^{\chi-1}\right)du\right]C_{\alpha,\gamma}(\tau)d\tau\Biggr).
\end{align*}On the other hand, if $\chi\in (1, 3/2)$, using
\begin{align*}
(u-\tau)^{\chi-1}=u^{\chi-1}-(\chi-1)u^{\chi-2}\tau
-\left[u^{\chi-1}-(\chi-1)u^{\chi-2}\tau-(u-\tau)^{\chi-1}\right],
\end{align*}
we have
\begin{align*}
\int_\tau^t u^{\chi-1}(u-\tau)^{\chi-1}du=&
\frac{t^{2\chi-1}}{2\chi-1}-\frac{\tau}{2}t^{2\chi-2}
-\frac{3-2\chi}{2(2\chi-1)}\tau^{2\chi-1}\\&-\int_\tau^t
u^{\chi-1}\left[u^{\chi-1}-(\chi-1)u^{\chi-2}\tau-(u-\tau)^{\chi-1}\right]du.
\end{align*}
Therefore, if $\chi\in (1, 3/2)$
\begin{align*}
\left\langle
[Y_{\alpha,\gamma}(t)]^2\right\rangle=&\frac{2}{\Gamma(\chi)^2}\Biggl(
\frac{t^{2\chi-1}}{2\chi-1}\int_0^t
C_{\alpha,\gamma}(\tau)d\tau-\frac{t^{2\chi-2}}{2}\int_0^t \tau
C_{\alpha,\gamma}(\tau)d\tau \\&-\frac{3-2\chi}{2(2\chi-1)}\int_0^t
\tau^{2\chi-1}C_{\alpha,\gamma}(\tau) d\tau
\\&-\int_0^t \left\{\int_\tau^t
u^{\chi-1}\left[u^{\chi-1}-(\chi-1)u^{\chi-2}\tau-(u-\tau)^{\chi-1}\right]du\right\}C_{\alpha,\gamma}(\tau)d\tau\Biggr).
\end{align*}The large--$t$ behaviors of $\int_0^{t}C_{\alpha,\gamma}(\tau)
d\tau$ and $\int_0^t \tau C_{\alpha,\gamma}(\tau)d\tau$ have been
studied and given by \eqref{eq4_17_14}, \eqref{eq4_17_11},
\eqref{eq4_17_12} and \eqref{eq4_17_13}. For the term $\int_0^t
\tau^{2\chi-1}C_{\alpha,\gamma}(\tau)d\tau$, since
$C_{\alpha,\gamma}(\tau)(\tau)\sim
A_1\tau^{-2\alpha-1}+O(\tau^{-4\alpha-1})$ as $\tau\rightarrow
\infty$, we find that if $2\chi-1< 2\alpha$,
\begin{align*}
\int_0^{t} \tau^{2\chi-1} C_{\alpha,\gamma}(\tau) d\tau
=&\int_0^{\infty} \tau^{2\chi-1} C_{\alpha,\gamma}(\tau) d\tau
-\int_t^{\infty} \tau^{2\chi-1}
C_{\alpha,\gamma}(\tau)d\tau\\=&B_\chi + O(t^{-2\alpha+2\chi-1}).
\end{align*}However, if $2\chi-1>2\alpha$,
\begin{align*}
\int_0^t \tau^{2\chi-1} C_{\alpha,\gamma}(\tau)d\tau = B_\chi
t^{2\chi-2\alpha-1}+O(t^{\max\{0, 2\chi-1-4\alpha\}});
\end{align*}and if $2\chi-1=2\alpha$, then
\begin{align*}
\int_0^t \tau^{2\chi-1} C_{\alpha,\gamma}(\tau) d\tau = B_\chi\log
t+O(1).
\end{align*}Now for the term
\begin{align*}
\int_0^t \left[\int_\tau^t
u^{\chi-1}\left(u^{\chi-1}-(u-\tau)^{\chi-1}\right)du\right]C_{\alpha,\gamma}(\tau)d\tau,
\end{align*}By  making a change of variable $u\mapsto u\tau$, we have
\begin{align*}
&\int_0^t \left[\int_\tau^t
u^{\chi-1}\left(u^{\chi-1}-(u-\tau)^{\chi-1}\right)du\right]C_{\alpha,\gamma}(\tau)d\tau\\
=&\int_0^t \tau^{2\chi-1}\left[\int_1^{\frac{t}{\tau}}
u^{\chi-1}\left(u^{\chi-1}-(u-1)^{\chi-1}\right)du\right]C_{\alpha,\gamma}(\tau)d\tau\\
=& \int_1^{\infty}u^{\chi-1}\left(u^{\chi-1}-(u-1)^{\chi-1}\right)
\int_0^{\frac{t}{u}}\tau^{2\chi-1}C_{\alpha,\gamma}(\tau)d\tau du.
\end{align*}Using the fact that $C_{\alpha,\gamma}(\tau)>0$, we find that this term is
bounded above by
\begin{align*}
\int_1^{\infty}u^{\chi-1}\left(u^{\chi-1}-(u-1)^{\chi-1}\right)du
\int_0^t \tau^{2\chi-1}C_{\alpha,\gamma}(\tau)(\tau)d\tau.
\end{align*}Notice that $\chi\in (1/2,1)$ implies that the first
integral is convergent. Similarly, we have for $\chi\in (1, 3/2)$,
\begin{align*}
&\int_0^t \left\{\int_\tau^t
u^{\chi-1}\left[u^{\chi-1}-(\chi-1)u^{\chi-2}\tau-(u-\tau)^{\chi-1}\right]du\right\}C_{\alpha,\gamma}(\tau)d\tau
\\\leq & \int_1^{\infty} u^{\chi-1}\left[u^{\chi-1}-(\chi-1)
u^{\chi-2}-(u-1)^{\chi-1}\right]du \int_0^{t}
\tau^{2\chi-1}C_{\alpha,\gamma}(\tau)d\tau;
\end{align*}and $\chi\in (1, 3/2)$ guarantees the convergence of the first integral.
Gathering the results, we find that when $t\rightarrow
\infty$,
\begin{align*}
\left\langle\left[Y_{\alpha,\gamma}(t)\right]^2\right\rangle\sim
\mathcal{A} t^{2\chi-1}+O\left( \max\{t^{2\chi-2\alpha-1},
t^{2\chi-2}, t^{2\chi-2}\log t, t^{0}, t^{0}\log t\}\right),
\end{align*}where
\begin{align*}
\mathcal{A}=\frac{2}{(2\chi-1)\Gamma(\chi)^2}\int_0^{\infty}
C_{\alpha,\gamma}(\tau)
d\tau=\frac{2B\lambda^{-2\alpha\gamma}}{(2\chi-1)\Gamma(\chi)^2},
\end{align*}and \eqref{eq4_16_7_1} follows.

\end{document}